# Particle diffusion in extracellular hydrogels

Federica Burla[1], Tatjana Sentjabrskaja[1], Galja Pletikapic[1], Joey van Beugen[1], Gijsje H. Koenderink[1,2,#]

[1] AMOLF, Department of Living Matter, Biological Soft Matter group, Science Park 104, 1098 XG Amsterdam, the Netherlands

[2] Department of Bionanoscience, Kavli Institute of Nanoscience Delft, Delft University of Technology, Van der Maasweg 9, 2629 HZ Delft, the Netherlands

[#] Corresponding author: g.h.koenderink@tudelft.nl

**Hyaluronic acid is an abundant polyelectrolyte in the human body that forms extracellular hydrogels in connective tissues. It is essential for regulating tissue biomechanics and cell-cell communication, yet hyaluronan overexpression is associated with pathological situations such as cancer and multiple sclerosis. Due to its enormous molecular weight (in the range of millions of Daltons), accumulation of hyaluronan hinders transport of macromolecules including nutrients and growth factors through tissues and also hampers drug delivery. However, the exact contribution of hyaluronan to tissue penetrability is poorly understood due to the complex structure and molecular composition of tissues. Here we reconstitute biomimetic hyaluronan gels and systematically investigate the effects of gel composition and crosslinking on the diffusion of microscopic tracer particles. We combine ensemble-averaged measurements via differential dynamic microscopy with single-particle tracking. We show that the particle diffusivity depends on the particle size relative to the network pore size and also on the stress relaxation dynamics of the network. We furthermore show that addition of collagen, the other major biopolymer in tissues, causes the emergence of caged particle dynamics. Our findings are useful for understanding macromolecular transport in tissues and for designing biomimetic extracellular matrix hydrogels for drug delivery and tissue regeneration.**

## Introduction

Hyaluronan is a charged linear polysaccharide that is widely present in the human body, where it forms extracellular hydrogels[1]. Together with bottlebrush proteoglycans and polysialic acid, it forms a dense layer around cells known as the glycocalix, which protects cells against damage and virus infections. This slimy coating also modulates cell-cell and cell-matrix communication by affecting the mobility and accessibility of receptor proteins in the cell membrane[2,3]. Cancer cells have greatly upregulated levels of hyaluronan in their glycocalyx, which is thought to promote a tumor phenotype by increasing integrin adhesion and signalling[4–6]. In the cumulus cells that surround oocytes[7], hyaluronan forms a specialized jelly-like structure that is crosslinked by accessory proteins and is critical for fertilization[8,9].

In soft connective tissues, hyaluronan is not grafted to cells but organized in entangled or crosslinked hydrogels with the help of accessory proteins. In cartilage and in the vitreous humor, hyaluronan is for instance found together with collagen fibrils and aggrecans. The molecular weight, concentration, and network structure of hyaluronan in these extracellular hydrogels vary with tissue type and change during tissue development and with age[1,10]. These variations help tailor the biomechanical behavior of each tissue for its specific function and strongly influence cell physiology through mechanochemical signalling[11,12]. In addition, variations in the physical properties of hyaluronan hydrogels influence the ability of macromolecules (growth factors, nutrients, and signaling factors), virus particles and cells to diffuse or migrate through the tissue[13]. Many pathological situations are associated with excessive deposition of hyaluronan and changes in its molecular weight, with deleterious consequences for tissue penetrability[14]. For instance, excessive deposition of hyaluronan in demyelinated lesions during multiple sclerosis inhibits regrowth of nerves[15], and changes in hyaluronan content in malignant tissues hampers the transport and delivery of cancer therapeutics[16,17]. It is therefore important to understand how normal and pathological variations in hyaluronan molecular weight, concentration, and interactions with other matrix components impact the penetrability of the extracellular matrix. However, it is challenging to dissect the contribution of hyaluronan due to the complex composition and structure of living tissues. A way to overcome this difficulty is to reconstitute extracellular hydrogels from purified hyaluronan. Several previous studies have indeed reported measurements of particle mobility in reconstituted hyaluronan gels based on either passive microrheology (using video tracking or light scattering to detect the thermal motions of tracer particles[18–20]) or active microrheology (using optical tweezer manipulation of tracer particles[21]). With few exceptions[22-23], the focus of these studies was on single-component networks of hyaluronan networks that interact solely through excluded volume and electrostatic interactions.

Here we investigate the dynamics of both semidilute solutions and crosslinked gels of high molecular weight hyaluronan using a variant of passive microrheology known as differential dynamic microscopy (DDM)[24]. Rather than tracking the trajectories of individual tracer particles, this method probes the ensemble-averaged dynamics of many particles at once using a Fourier-space analysis of time-lapse movies. DDM analysis thus yields information comparable to data obtained by dynamic light scattering. DDM has a number of advantages over conventional particle tracking: it can be performed over a wider particle size range, can work with fluorescence, bright field and dark field microscopy data, and does not need complex tracking algorithms nor a costly setup[24,25]. For these reasons, DDM was recently introduced as a new method for performing microrheology on soft matter systems[26–28]. We complement the ensemble-averaged information obtained from DDM with single particle tracking, using the same time-lapse imaging data as input, in order to test for the presence of spatial and temporal heterogeneities in particle dynamics that typically arise in heterogeneous systems such as polymer gels[29–32]. In order to understand how the dynamics of hyaluronan networks are modulated by accessory extracellular molecules that introduce crosslinks, we probe single-component gels with three different crosslink configurations: semidilute solutions, transiently crosslinked gels obtained by pH-triggered gelation[33], and chemically crosslinked gels. We show that semidilute solutions and transiently crosslinked hyaluronan networks simply hinder particle transport through enhanced viscous drag, whereas permanently crosslinked gels hamper particle diffusion by size exclusion. We also study two-component gels combining hyaluronan with a fibrillar collagen network, and show that the composite polymer gels exhibit caged dynamics. Our data reveal a rich phase-space for control over particle diffusion in the crowded extracellular space of tissue, both through variations in the physical properties of hyaluronan itself and through interactions with other matrix constituents. Our results can be useful in interpreting the impact of hyaluronan on cell-cell signaling[13] and on drug penetration through the interstitial matrix of tissues to solid tumours[17], as well as for the design of hyaluronan-based hydrogels for regenerative medicine and controlled drug release[34].

## Materials and methods

### Bead passivation protocol

Polystyrene tracer particles with diameters of 0.6 µm as specified by the supplier were purchased from Sigma Aldrich. Red fluorescent particles of 0.2 or 0.1 µm diameter as specified by the supplier were purchased from ThermoFischer Scientific (Fluoro-Max Dyed Red Aqueous Fluorescent Particles). To prevent non-specific interactions of the particles with the hyaluronan polymers, the particles were passivated with poly(ethylene glycol) chains following an established protocol[35]. Briefly, 45 µL of a stock solution of particles was pipetted in an Eppendorf tube and sonicated for 5 minutes to disperse any aggregates. Next, 300 µL Pluronics F-128 (10% w/v, Sigma Aldrich) was added and incubated with the particles for 10 minutes. Subsequently, 120 µL toluene (99.5%, Sigma Aldrich) was added and the samples were incubated on a rotating wheel (18 rpm) for three hours. Toluene swells the particles, allowing for the hydrophobic blocks of the Pluronic chains to insert into the particle surface. Afterwards, toluene was removed by heating the Eppendorf tubes under the fume hood at 95°C for 15 minutes. The particles were centrifuged 5 times (10 minutes at 5000 rpm each time) with replacement of the supernatant by MilliQ water between each centrifugation step. This washing procedure deswells the particles, so the Pluronic chains become firmly anchored to the particle surface. The passivated particles were stored in the fridge for a maximum of 1 month. Before use, the particles were sonicated for 15 minutes to remove any aggregates. The hydrodynamic diameters of the particles, as determined from DDM measurements on dilute particle suspensions in water, were 0.15 µm, 0.22 µm and 0.66 µm, which is larger than the nominal radii likely due to the combined presence of the Pluronic layer and a solvation shell.

### Image acquisition

Time-lapse videos used in both single particle tracking and differential dynamic microscopy analysis were acquired on an inverted Ti-Eclipse microscope (Nikon) with a 100x oil or 60x water immersion objective (Nikon) with numerical apertures of 1.49 and 1.27, respectively, and with a digital CMOS camera (Hamamatsu, Orca-Flash 4.0). For the larger 0.6 µm particles, we imaged with bright field mode using illumination with white Halogen light (Nikon). The smaller 0.2 and 0.1 µm particles were imaged with fluorescence microscopy, using a 532 nm laser (Lumencor LED) for illumination. The exposure time of the camera was set to 10 ms, giving access to a 100 fps acquisition rate. The delay time between each frame was varied from 0 to 100 ms depending on the sample viscosity, in order to obtain a full relaxation of the intermediate scattering function (or at least a partial relaxation in the case of chemically crosslinked networks). Each video consisted of 5000 frames and for each condition reported, videos were recorded from at least three different regions of the same sample and in at least three independently prepared samples. Collagen-hyaluronan composites were imaged in confocal reflectance mode, a method which allows label-free imaging, using a 488 nm Argon laser (Melles Griot) for illumination. Images were taken from 10 µm above the surface over a depth of 30 µm, with a step size of 0.5 µm, and are shown as maximum projection intensity.

### DDM analysis

DDM analysis of the time lapse videos was performed using a custom written MatLab program based on principles explained in prior studies[24]. The thermal motions of the tracer particles in the viscoelastic medium provided by the hyaluronan hydrogels cause temporal fluctuations of the intensity in each image pixel with coordinates ($x,y$). Fourier transformation of the intensity time traces $I(x,y,t)$ and correlation of the difference images at different lag times $\Delta t$ produces the image structure function $D(\boldsymbol{q},\Delta t)$:

$$D(\boldsymbol{q},\Delta t) = <|i(\boldsymbol{q},t+\Delta t) - i(\boldsymbol{q},t)|^2> \qquad (1)$$

where $q = 2\pi/l$ is the wavevector and $l$ represents a characteristic distance in real space. The intermediate scattering function $f(q,\Delta t)$ is obtained from the image structure function through the formula:

$$D(\boldsymbol{q},\Delta t) = A(q)[1 - f(\boldsymbol{q},\Delta t)] + B(\boldsymbol{q}), \qquad (2)$$

where $A(q)$ is a proportionality factor and $B(q)$ represents the noise of the camera. By fitting $f(q,\Delta t)$ to a phenomenological stretched exponential decay, $exp\{-(\frac{\Delta t}{\tau(q)})^n\}$ with stretching exponent $n$, one retrieves the diffusion coefficient of the tracer particles, $D = 1/(\tau(q)q^2)$. In a Newtonian fluid of viscosity $\eta$, the diffusion coefficient of particle of radius $r$ is inversely proportional to the viscosity according to the Stokes-Einstein relation:

$$D_m = \frac{k_B T}{6\pi\eta r} \qquad (3)$$

Under certain conditions, this relation can be generalized to complex fluids that are viscoelastic:

$$\tilde{G}(s) = \frac{k_B T}{\pi a s <\widetilde{r^2}(s)>}, \qquad (4)$$

where $\tilde{G}(s)$ and $<\widetilde{r^2}(s)>$ represent the Laplace transforms of, respectively, the complex viscoelastic modulus of the fluid and the mean-squared displacement of the particles. This generalization is the basis for passive microrheology, a technique that since its inception[36] has been widely applied to study polymers, including biopolymers such as hyaluronan[18,21]. One important condition for the generalized Stokes-Einstein relation to hold is that the tracer particles are large enough compared to the correlation length of the polymer network such that they perceive the network as a viscoelastic continuum[37]. In entangled polymer solutions, the tracer particle motions will then reflect the bulk solution viscosity. In order to extract the frequency-dependent storage modulus G' and loss modulus G'', we first transformed the ISF at q = 4.5 $\mu m^{-1}$ to obtain the mean-squared displacement according to the following relation[27] valid for displacements in 2D:

$$<\Delta r^2(\Delta t)> = -\frac{4}{q^2} \ln(f(q,\Delta t)) \qquad (5)$$

We subsequently employed the fitting routine implemented in ref.[38] that is based on the Evans-Tassieri method[39].

### Particle tracking analysis
Single particle centroid tracking was performed with TrackPy[40], an algorithm based on the Crocker-Weeks tracking algorithm[41]. This algorithm allows to track and drift-subtract individual particle trajectories and to retrieve the corresponding mean-squared-displacements. We next reloaded the trajectories to a custom-written Python program to calculate the ensemble-averaged van Hove distribution functions of particle displacements and to perform curve fitting with Gaussian and exponential functions. For each video, we analyzed a total of around 1000 tracks, and for each sample we analyzed at least three independently prepared samples. We restricted the analysis to lag times Δt corresponding to at most 10 % of the total length of the video, in order to avoid artefacts deriving from low statistics.

### Sample preparation
Semidilute solutions of hyaluronic acid were prepared by dissolving sodium hyaluronate obtained by fermentation of *Streptococcus Equii* bacteria with a nominal molecular weight between 1.1-1.6 MDa (Sigma Aldrich) at a concentration of 1,2 or 4 mg/mL in phosphate-buffered saline (PBS: pH 7.4, 138 mM NaCl; 2.7 mM KCl) obtained in tablet form from Sigma Aldrich. The samples were vigorously vortexed for several minutes or placed on a spinning wheel at room temperature for several hours to ensure full solubilisation. Transiently crosslinked hyaluronan gels at pH 2.5 were obtained by adding varying amounts of an aqueous solution of 15 mM HCl to the semidilute solutions of hyaluronan. Chemically (i.e. permanently) crosslinked hyaluronan solutions were obtained by combining commercially available thiolated hyaluronic acid (Glycosil, 2B Scientific, 240 kDa) in powder form (2B Scientific) with a poly(ethylene glycol) diacrylate crosslink (Extralink PEGDA, 2B Scientific, 3.4 kDa) in PBS buffer. Composite hyaluronan-collagen samples were prepared by neutralizing type I collagen with intact telopeptide end-sequences from bovine dermis (TeloCol, CellSystems, supplied at 3.0 mg/mL in hydrochloric acid) by the addition of 1 M NaOH and PBS from a 10x stock concentration for buffering and quickly mixing this on ice with 2 mg/ml hyaluronan before collagen started to polymerize. Collagen was allowed to polymerize at 37°C for two hours before imaging. Passivated tracer particles were added directly before the measurements in a ratio of 1:100 with the final sample volume, and the solution was homogenized by vigorous vortexing. In the case of crosslinked hyaluronan and composite collagen/hyaluronan samples, the particles were added prior to hyaluronan crosslinking or collagen polymerization, in order to ensure a homogeneous distribution across the sample.

### Shear rheology

The linear viscoelastic moduli of semidilute and crosslinked hyaluronan networks were measured by small amplitude oscillatory shear rheology on a stress-controlled rheometer (Anton Paar MCR 501) equipped with a cone-plate geometry with a diameter of 40 mm, cone angle of 1°. The experiments were performed at a temperature of 22°C set by a Peltier system. Semidilute hyaluronan solutions were loaded onto the bottom plate with a pipette, putty samples were loaded with a spatula, and permanently crosslinked samples were quickly loaded with a pipette before crosslinking set in. The samples were allowed to thermally equilibrate for 10 minutes in the case of the semidilute and putty solutions, while the crosslinked gels were allowed to polymerize for two hours between the rheometer plates before measuring. After equilibration, we we determined the linear viscoelastic moduli by applying an oscillatory shear strain at different oscillation frequencies, logarithmically spaced between 0.1 and 10 Hz, and a small strain amplitude of 0.5%. To determine the viscosity, we furthermore measured flow curves by applying a rotational shear with the strain rate increasing logarithmically from 0.01 to 100 s$^{-1}$. The reported results are averages of at least three independent measurements for each sample condition. The measurements on collagen and collagen-hyaluronan composites were performed after allowing for *in situ* collagen polymerization for two hours.

### Mesh size determination

We inferred the mesh size of the crosslinked hyaluronan gels from the measured linear elastic shear modulus $G'$ by referring to rubber elasticity theory[42,43], which predicts:

$$G' = \rho_{el} k_B T, \tag{6}$$

where $k_BT$ is the thermal energy and $\rho_{el}$ represents the number density of elastically active network strands, related to the mesh size $\varphi_{el}$ by:

$$\varphi_{el} = \rho_{el}^{-1/3} \tag{7}$$

We thus estimate a mesh size $\varphi_{el}$ = 200 nm for the 4 mg/mL crosslinked hyaluronan gels, for which we measured $G'$ = 4 Pa. For the calculation of the mesh size of uncrosslinked hyaluronan, we employed the fact that the mesh size of a solution of worm-like chains polymer interacting through excluded volume only should scale according to[44]:

$$\varphi = \sqrt{3/\nu L}, \tag{8}$$

where $L$ is the polymer length, and $\nu$ is the number of polymers per unit volume. We find values around 200 nm for the 4 mg/mL solution, comparable to the mesh size of the crosslinked network. This observation indicates that the network remains homogeneous in the presence of crosslinking.

To determine the mesh size of the 1 mg/ml collagen networks, we used a previously reported image analysis algorithm[45]. Briefly, we binarized a 3D-confocal image stack and measured the distances between on and off pixels. The distance distribution was well-fitted by an exponential decay, and the decay exponent was taken as the average mesh size after conversion from pixels to µm (121 nm/pixel). For 2 mg/ml collagen networks, we calculated the mesh size assuming that it scales with concentration according to $c^{-1/2}$. We obtained average mesh size values of 3 µm for 1 mg/mL collagen and 2 µm for 2 mg/mL collagen networks.

The mesh size of the composite networks was calculated as a geometrical average of the mesh sizes of the hyaluronan and collagen networks [ref.[46]]:

$$\varphi_c^{-3} = (\varphi_{HA}^{-3} + \varphi_{coll}^{-3}), \qquad (9)$$

where $\varphi_c$ is the composite mesh size, $\varphi_{HA}$ is the hyaluronan mesh size, $\varphi_{coll}$ is the collagen mesh size. We obtained a mesh size of 300 nm for a composite network composed of 1 mg/mL collagen and 2 mg/mL hyaluronan, almost identical to the mesh size of the hyaluronan-only network, consistent with the large mismatch between the hyaluronan and collagen mesh sizes.

## Results

### Effect of crosslinking on particle diffusivity in hyaluronan networks

We first measured particle diffusion in hyaluronan networks at different concentrations ranging from 1 to 4 mg/mL as a function of the state of chain crosslinking (Figure 1(a-c)). The overlap concentration $c^*$ for hyaluronan with a molecular weight of 1 MDa is around 2 mg/mL[18,19]. Given that entanglements typically set in at an entanglement concentration $c_e$ that is at least 5 times larger than $c^*$, the solutions should fall in the semidilute unentangled regime.[47] We verified this by measuring the concentration-dependence of the viscosity, which in the semidilute unentangled regime[48] scales with a power law of around 1.3 (see Supplementary Figure S1). *Semidilute solutions* were obtained by dissolving hyaluronan in physiological salt buffer (PBS) at neutral pH, where the electrostatic charges are screened and the polymers behave as random coils that interact mainly via excluded volume and hydrodynamic interactions[49]. *Transiently crosslinked networks* were obtained by lowering the pH to 2.5, which reduces electrostatic repulsions among hyaluronan chains due to the proximity to its isoelectric point of 2.5 [ref. [50]] and enhances chain associations through hydrogen bonds between amide and carboxylate residues[33]. This pH-induced gel state is traditionally referred to in the hyaluronan literature as the *putty* state[51]. Finally, *permanently crosslinked networks* were obtained by reacting thiol-modified hyaluronic acid with a diacrylate (PEGDA) crosslinker. We note that the measurements reported with the chemically crosslinked hyaluronan refer to a lower molecular weight than the entangled and transiently crosslinked hyaluronan samples. As shown in Supplementary Figure S2, DDM reveals similar dynamics for particles in solutions of the low (240 kDa) and high (1.5 MDa) molecular weight semidilute hyaluronan, but with a higher particle mobility, reflecting the lower solution viscosity associated with the lower molecular weight.

Particle diffusion in extracellular hydrogels

We seeded the hyaluronan networks with tracer particles with a diameter of 0.6 μm and measured the ensemble-averaged dynamics of these particles by recording time-lapse movies and performing DDM analysis. As shown in Figure 1, the dynamics of the particles as quantified through the intermediate scattering function (ISF) strongly varied with crosslinking conditions. Note that the scattering functions are shown for a *q*-value of 4.65 μm$^{-1}$, which is an intermediate value where the ISF is not affected by the drifting of the particles out of the field of view at high *q* (see ref.[52] and Supplementary Figure S3). Particles in semidilute solutions of hyaluronan (Fig. 1a) exhibited simple diffusive behavior characterized by a single-exponential decay of the intermediate scattering function in the concentration range here investigated (Fig. 1d). A similar decay was observed in different regions of interest in a given sample, indicating that the material was homogeneous over length scales of hundreds of micrometers (the size of the field of view of the microscope) (see Supplementary Figure S4). Also, we observed comparable decays for different samples. The observations of simple diffusion and negligible spatial heterogeneity are consistent with the macroscopic rheology of the hyaluronan solutions, which behaved as viscous solutions with negligible elasticity (see Supplementary Figure S5a). Particles in the transiently crosslinked network obtained by pH-induced gelation (Fig. 1b) showed slower dynamics, although the intermediate scattering function still exhibited a single-exponential decay (Fig. 1e). This behavior is again qualitatively consistent with the macroscopic rheology of the gels, which behaved as viscoelastic Maxwell fluids (see Supplementary Figure S5b).

By contrast, particles in the permanently crosslinked network (Fig. 1c) showed different dynamics, depending on the concentration (Fig. 1f). Particles in gels made of 1 mg/ml hyaluronan showed simple diffusion with a diffusion coefficient slightly lower than in water. Particles in 2 mg/ml hyaluronan gels showed incomplete relaxation of the intermediate scattering function and particles in 4 mg/mL gels did not measurably diffuse at all. Indeed, the 4 mg/ml gels behaved rheologically as elastic solids with a frequency-independent elastic modulus that was much larger than the loss modulus (see Supplementary Figure S5c). At concentrations lower than 3 mg/mL, as we previously reported[53], the samples were too soft to reliably measure an elastic modulus. We note that the ISFs for 4 mg/mL and 1 mg/mL hyaluronan networks were homogeneous across different fields of view, whereas the ISFs at 2 mg/mL showed substantial heterogeneity (see Supplementary Figure S4).

To quantify the particle dynamics, we fitted each intermediate scattering function (at q=4.65 μm$^{-1}$) to a stretched exponential decay $e^{-(\frac{\Delta t}{\tau(q)})^n}$, allowing us to extract the transport coefficient $D = 1/(\tau(q)q^2)$ and the stretching exponent *n*. We furthermore compared the exponents *n* with the subdiffusive exponent α obtained from fitting the decay of τ(q) (over the range q = 1-10 μm$^{-1}$) to the form:

$$\tau(q) = (Kq^2)^{-1/\alpha}, \tag{10}$$

to recover the transport coefficient *K* and subdiffusive exponent α. For the semidilute solutions and the transiently crosslinked networks obtained by pH-induced gelation, we found stretching exponents *n* close to 1 (see Supplementary Figure S6), consistent with simple diffusion. Consistent with this, also the subdiffusive exponent α from fitting the decay of τ(q) was close to 1 (see Supplementary Figure S6 and S7). The subdiffusive exponent for the mean-squared displacement $<\Delta r^2> = 2K(\Delta t)^\alpha$ measured by particle tracking analysis was also close to 1, although we notice that the temporal range of the MSD was more limited than that of the DDM data due to low statistics at long lag times and limitations posed by the tracking accuracy at short lag times[27]. For the chemically crosslinked gels, we could only reliably determine α for 1 mg/ml, which was equal to one, because of the incomplete relaxation of the ISF at 2 and 4 mg/mL networks. The incomplete relaxation of the ISF suggests that for these two cases the particle motion is subdiffusive. In case of the 2 mg/mL system, this is accompanied by high spatial heterogeneity of the DDM correlation functions, as mentioned earlier. Similar behavior was recently reported for actin-microtubule composite networks, where subdiffusion was also accompanied by large heterogeneity[54].

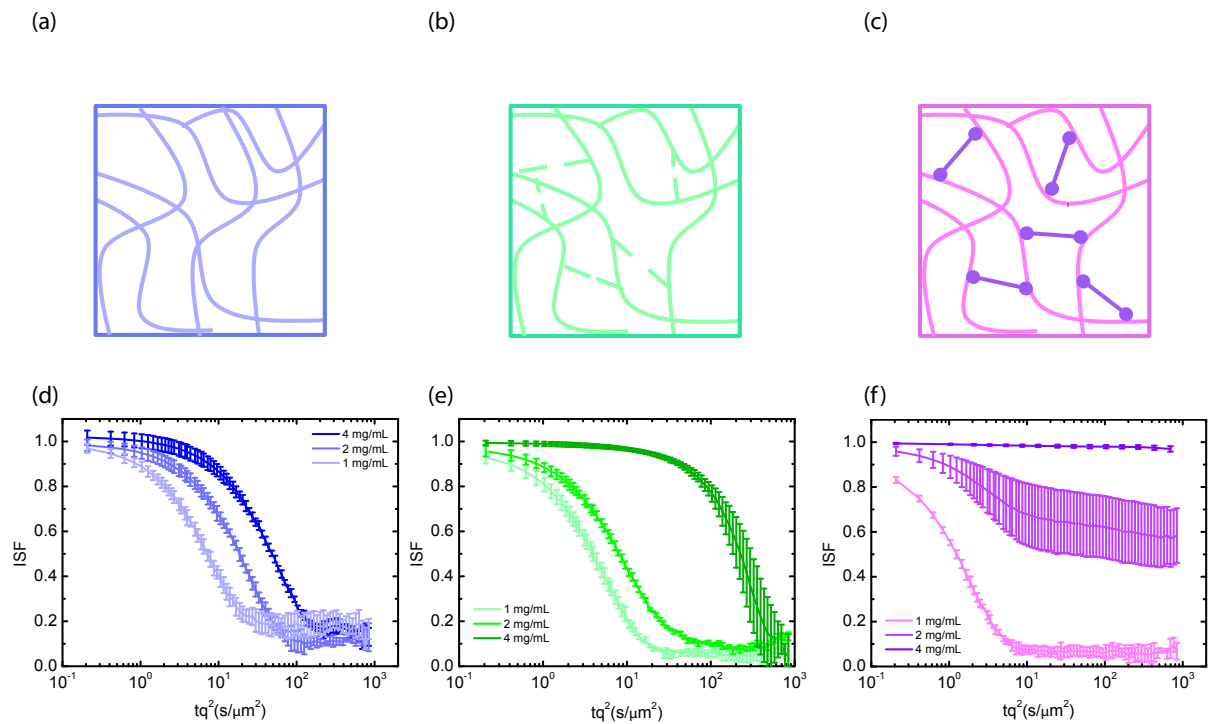

**Figure 1: Diffusion of 0.6 µm tracer particles in hyaluronan networks with various degree of crosslinking.** (a-c) Schematic overview of the different configurations of hyaluronan that were investigated: semidiluted (a), transiently crosslinked by pH-induced hydrogen-bonding ('putty', b), and chemically crosslinked by short PEGDA polymer links ('crosslinked', c). (d-f) Corresponding intermediate scattering functions (ISF) at different concentrations of hyaluronan (see legends), averaged over at least three measurements per condition. Time on the x-axis is multiplied by $q^2$, where q = 4.65 µm$^{-1}$.

The ensemble-averaged DDM analysis suggests that semidilute hyaluronan solutions and transient gels formed by acid-induced gelation behave as simple viscous media. Microrheology analysis of the data for entangled and transient hyaluronan networks confirm that these samples behave as viscoelastic fluids and furthermore show that the apparent viscosity experienced by the probe particles is about one order of magnitude lower than the macroscopic viscosity measured by rheology (see Supplementary Figure S8). This discrepancy is consistent with other microrheology data on polymer solutions, where this effect has been ascribed to polymer depletion from near the probe particles[37]. By contrast, the ensemble-averaged DDM analysis suggests that particles in chemically crosslinked hyaluronan networks are hindered by size exclusion since particles get increasingly immobilized with increasing polymer concentration (and hence decreasing pore size).

To test the dependence of particle mobility in the crosslinked gels on the ratio between particle size and pore size, we measured the diffusivity of smaller tracer particles with diameters of 0.1 and 0.2 µm. At 1 mg/mL, the particles experienced little hindrance from the network, irrespective of their size (Fig. 2a). The diffusion coefficient was in fact close to that in solvent and the stretching exponent $n$ was close to 1. Furthermore, the ISFs collapsed onto a single master curve upon rescaling the time axis with the tracer diameter, consistent with the Stokes-Einstein relation (see Supplementary Figure S9). By contrast, at 2 mg/mL, the particle mobility did depend on the particle size: the smallest (0.1 µm) particles still diffused freely, while 0.2 µm and 0.6 µm particles moved subdiffusively, as evidenced from incomplete relaxation of the ISF (Fig. 2b) and from subdiffusive exponents $\alpha$ smaller than 1 in case of the 0.2 µm partices (see Supplementary Figure S10). The ISFs did not collapse onto a single curve upon rescaling the time axis with the tracer size. This observation demonstrates that the generalized Stokes-Einstein (GSE) relation, which would predict an inverse dependence of the relaxation time on particle size, does not hold. We indeed expect a breakdown of the GSE relation, because the dependence of particle mobility on the ratio between probe size and network mesh size indicates that size exclusion governs the mobility rather than the macroscopic viscosity (see Supplementary Figure S9b). In 4 mg/mL hyaluronan gels, the smallest (0.1 µm) particles still showed limited mobility, while the two larger particles (0.2 µm and 0.6 µm) were immobilized. Again, the ISFs did not collapse onto a single curve upon rescaling the time axis with the tracer size (see Supplementary Figure S8c), demonstrating a breakdown of the GSE relation.

The onset of immobilization which we observe here is consistent with independent estimates of the ratio between particle radius and mesh size. For the 4 mg/mL crosslinked networks, we estimate an average mesh size of 200 nm from the network elastic modulus. This value is consistent with the observation that the particles of diameter 0.2 µm are stuck in the network, while the 0.1 µm particles can move albeit being slowed down. At 2 mg/mL, we estimate from the scaling of mesh size with concentration an average mesh size of around 300 nm. However, in this case previous macroscopic rheology data indicated that the network is not fully percolated[53]. This likely explains why we observe heterogeneous dynamics for the 0.6 µm particles with large differences among ISFs measured in different sample regions (see Supplementary Figure S4). Finally, at 1 mg/mL, previously reported rheology data indicated that the network is not percolated[53], and indeed the particles experience free diffusion, governed by an apparent viscosity only slightly smaller than the solvent viscosity (see Supplementary Information S10).

Particle diffusion in extracellular hydrogels

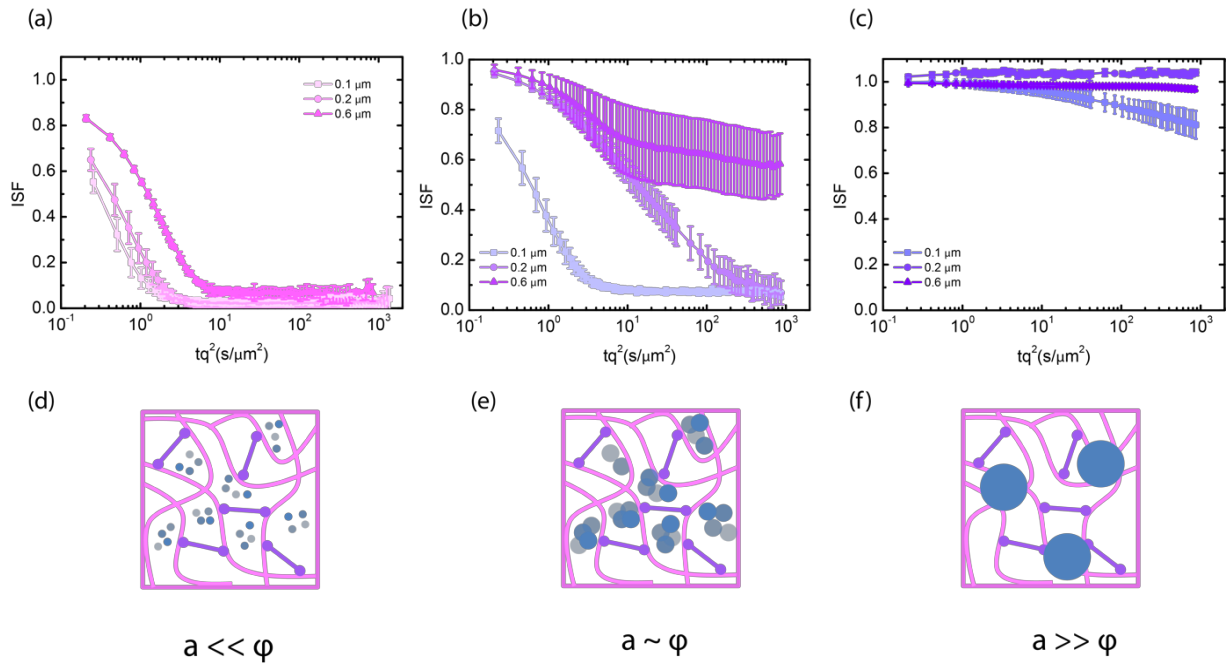

Figure 2: Mesh size effects on tracer particle mobility in chemically crosslinked hyaluronan networks. (a-c) ISF for three different tracer particle sizes (legends) in crosslinked gels with hyaluronan concentrations of 1 mg/mL (a), 2 mg/mL (b) and 4 mg/mL (c). (d-f) Schematic interpretation. When particle size $a$ is much smaller than the network mesh size, they effectively experience the solvent viscosity with little hindrance from the network (d). (e) Particles with size $a$ comparable to the mesh size experience hindered diffusion. (f) Particles with size $a$ larger than the mesh size are immobilized.

A microrheology analysis for the 4 mg/mL permanently crosslinked gel shows that particles of 0.6 μm size report an elastic plateau modulus comparable with the modulus measured by rheology, suggesting that the particles feel the elastic response of the networks (see Supplementary Figure S8). We notice a dependence on the size of the particles used for microrheology analysis, consistent with the presence of structures in the network with characteristic sizes larger than the particles' size[55].

To test whether the densest (4 mg/mL) hyaluronan networks featured any spatial or temporal heterogeneities at the level of single-particle trajectories, we complemented the ensemble-averaged DDM analysis with an analysis of the individual particle trajectories in real space. We focused on the van Hove distribution, which yields the probability of finding a particle at a distance $\Delta x$ or $\Delta y$ after a time interval $\Delta t$ and has been shown to provide insight in heterogeneities[31,56–58]. We note that the probability of displacements in the $x$ and $y$ directions were isotropic (See Supplementary Figure S11). For a simple fluid, the van Hove probability distribution should be Gaussian, whereas for complex fluids with spatial and/or temporal heterogeneities it has non-Gaussian tails that are generally exponential[47,58]. The distributions for all of the hyaluronan networks looked in first instance Gaussian, with a width $\sigma$ that increased with time due to diffusion (see Figure 3). However, the rate at which the van Hove functions broadened was lower for the transiently and permanently crosslinked networks than for the semidilute solution (see Supplementary Figure S12), consistent with the macroscopic rheology of the crosslinked samples being characteristic of an elastic solid rather than a fluid. We notice also that the distribution presents tails, which deviate from a Gaussian distribution. To quantify this deviation, we use the non-Gaussian parameter[46,60,61]:

$$\xi = \frac{<\Delta x(\tau)^4>}{3<\Delta x(\tau)^2>^2} - 1 \qquad (11)$$

This parameter is 0 for a Gaussian distribution and 2 for an exponential distribution, but it can assume larger values when the distribution function has heavy tails[62]. We found that ξ, for lag times where the particle displacement was larger than the tracking accuracy (~0.1 µm), was around 1 for all three types of hyaluronan samples (see Supplementary Figure S12). This value, intermediate between that of a Gaussian and that of an exponential distribution, indeed suggests the presence of spatial heterogeneities. Although the reproducibility of the DDM measurements for different regions of interest demonstrates that the samples are uniform over length scales of hundreds of microns (see Supplementary Figure S4), the non-Gaussianity of the van Hove distribution reveals some heterogeneity of the material at the micron scale.

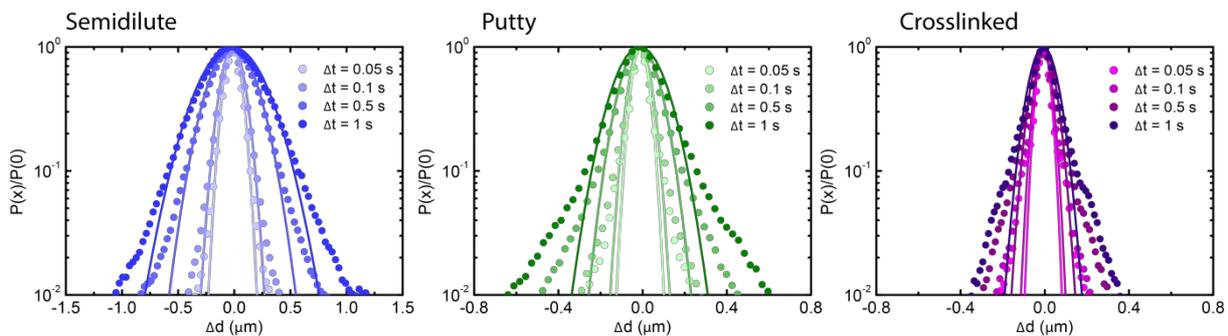

**Figure 3: Real-space analysis of the individual particle trajectories in hyaluronan networks for different states of crosslinking.** Gaussian fits (solid lines) to Van Hove distributions (squares) at different lag time intervals, ranging from 0.01 to 1 s (legends), for semidilute solutions (blue, left), transiently crosslinked (putty) networks at pH 2.5 (green, middle) and permanently crosslinked networks (purple, right). The time-dependence of the width of the van Hove functions and the non-Gaussian parameter is shown in Supplementary Figure S12.

## Composite hydrogels of hyaluronan and collagen

Since hyaluronan in many connective tissues is found in conjunction with fibrillar collagen, we then investigated the effect of interactions between hyaluronan and collagen on particle transport using DDM. We started from fibrillar collagen networks polymerized at 1 mg/mL and measured how the particle dynamics changed when hyaluronic acid was included, prior to polymerization, at a concentration of 2 mg/mL. As shown in Figure 4a, the intermediate scattering function for embedded 0.6 µm tracer particles revealed full relaxation in the two single-component cases of pure collagen (black) and pure hyaluronan (blue). Particle diffusion in collagen networks was only slightly slower than in pure solvent, likely because the mesh size of the network (~3 µm, Fig. 4b) was much larger than the particle size (see microrheology analysis in Supplementary Figure S13). The slightly enhanced drag on the particles could reflect the presence of a small fraction of non-polymerized collagen[63] or hydrodynamic drag imposed by the collagen fibrils. Surprisingly, particles in composite networks (orange) showed relaxation dynamics that were qualitatively different from the single-network responses: after a partial initial decay, the intermediate scattering function developed a plateau indicative of caged particle motions. The subdiffusive exponent α for the composite network as determined from fitting τ(q) was ~0.5 (see Supplementary Figure S14), indicating subdiffusive motion[57-64]. Through confocal imaging, we verified that the collagen network architecture was not affected by the presence of hyaluronan during polymerization (Fig. 4b-c). Qualitatively, neither the mesh size of the network nor the spatial organization of the fibers was significantly affected by the presence of hyaluronan. We quantitatively verified this by determining the mesh size of the collagen network within the composite network through image analysis. We found an average mesh size of 3.00 ± 0.05 µm in the presence of hyaluronan, which is indeed comparable to the mesh size of 3.1 ± 0.2 µm for the collagen-only network.

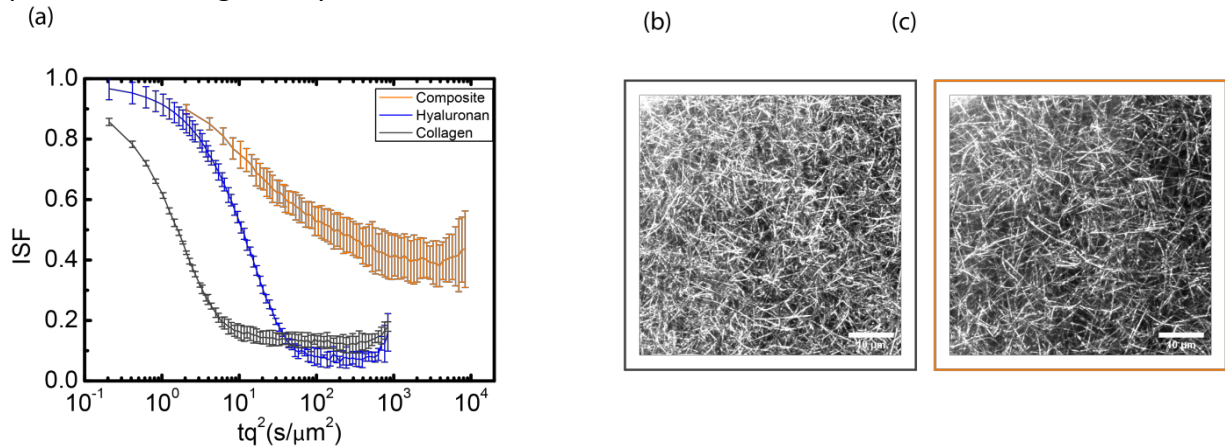

**Figure 4: Particle dynamics in composite collagen-hyaluronan networks.** (a) Intermediate scattering function (ISF) for 0.6 µm particles in networks of 1 mg/mL collagen (black curve), 2 mg/mL hyaluronan (blue) and composite collagen-hyaluronan (orange). (b-c) Confocal images of the fibrillar collagen network for a 1 mg/mL pure collagen network (b) and for a composite where the collagen fibrils are embedded in a hyaluronan background network that is not visible in the image (c). Scale bar indicates 10 µm.

To further test for caged dynamics in the composite networks, we analyzed the shape of the van Hove distribution functions constructed from the individual particle trajectories (see Figure 5). Similar to the pure hyaluronan systems, also the van Hove functions for the collagen network and the composite networks showed exponential tails, indicating microscale heterogeneities. A previous study[65] predicted that the exponential tails of the heterogeneous van Hove distribution should show a power-law scaling with time $\lambda = at^b$, with an exponent b of approximately 0.5, which results mathematically from a superposition of multiple Gaussians with different width (mirroring the heterogeneity of the sample). We observed values of $b$ close to 0.5 for the pure hyaluronan gels at 2 mg/mL and a lower value of 0.3 for the collagen sample and the composite network (see Supplementary Information S15), similar to a previous report on heterogeneous networks[56]. The non-Gaussian parameter $\xi$ characterizing the deviation of the van Hove distributions from a Gaussian reached values of 1 for 2 mg/mL collagen, 0.8 for 1 mg/mL collagen and 0.4 for hyaluronan at long lag times. For the composite networks, $\xi$ reached a value of 2 at long lag times, suggesting a higher degree of heterogeneity (see Supplementary Figure S16).

Particle diffusion in extracellular hydrogels

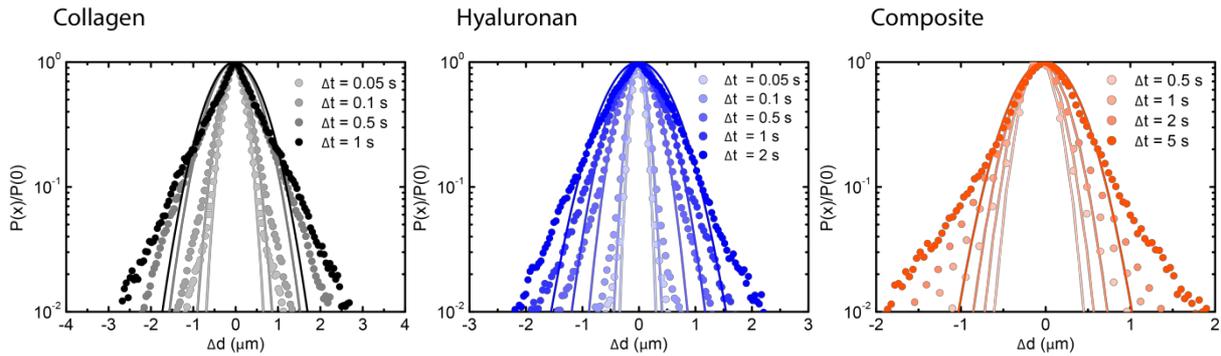

**Figure 5: Single particle analysis of tracer particle dynamics in collagen-hyaluronan composite networks.** Gaussian fits (solid line) to Van Hove distributions (squares) at different time intervals, ranging from 0.1 to 2 s, for collagen (gray, left), 2 mg/mL hyaluronan (blue, middle), and composite networks (orange, right).

To test whether subdiffusion results from polymer interactions or simply from an increased overall polymer density, we also performed measurements on collagen-only networks with a higher concentration of 2 mg/mL. In this case we observed incomplete relaxation of the ISF (see Supplementary information S17-S18), similar to the composite. The decay of τ(q) was characterized at small $q$ by a subdiffusive exponent α of 1, but showed deviations from this simple decay at larger values of $q$ (see Supplementary Information S14). Thus, subdiffusion was also observed for the collagen-only network upon raising the concentration, but to a lesser extent than in the collagen/hyaluronan composite. Since the mesh size of the 2 mg/ml collagen network is around 2 μm, much larger than the particle size, the caged dynamics we observe likely originates from the rigidity of the network and this can potentially also contribute to the caged dynamics in the composite network containing 1 mg/ml collagen. To test the effect of collagen on the rigidity of the composite system, we performed macroscopic rheology measurements. As shown in Supplementary Figure S19, the elastic modulus of the composite network was identical to that of a collagen-only network (1 mg/mL) network showing that collagen dominates the elastic response. This observation can be rationalized with the fact that hyaluronan by itself behaves as a viscous fluid with a negligible elastic response (see also Supplementary Figure S5). By contrast, the collagen network has a finite elasticity, consistent with predictions for crosslinked networks of athermal filaments[66,67].

## Discussion

Hyaluronan is a polyelectrolyte widely present in the extracellular matrix of connective tissues, where it regulates tissue biomechanics and cell physiology. In this study, we investigated how different degrees of hyaluronan crosslinking and combination with collagen fibrils, which are also an abundant component of tissues, influence the diffusion of tracer particles. To this end, we employed differential dynamic microscopy (DDM) and real space particle tracking. We first focused on the effect of the state of crosslinking by comparing particle dynamics in semidilute solutions and in transiently and permanently crosslinked networks. In semidilute solutions and transiently crosslinked networks of hyaluronan, the tracer particles exhibited simple diffusion characterized by a single exponential decay of the intermediate scattering function measured with DDM and by Gaussian statistics for the individual particle displacements. By contrast, in permanently crosslinked networks, the particle dynamics was strongly dependent on the particle size relative to the network pore size, with large particles being immobilized and small particles experiencing free diffusion. This size exclusion effect is reminiscent of similar effects observed in other gels in which diffusion was shown to be dominated by physical obstruction[68,69]. Particle motion in this case is strongly dominated by the ratio between the particle size and the characteristic pore size of the matrix. This finding is interesting when considering cases in which hyaluronan is cross-linked by binding proteins, as in the case of an anti-inflammatory response[70,71]. The induced crosslinking could, by modulating the network pore size, restrict the access of pathogenic factors by size-exclusion. The concept of size-filtering from the extracellular matrix was previously reported to be one of the main mechanisms by which the ECM acts as a diffusion barrier[72]. We emphasize, however, that here we focused on the diffusion of uncharged, sterically stabilized particles that do not interact with the matrix. More complex additional hindrance effects can arise when considering the diffusion of charged particles that interact with the extracellular matrix through electrostatic attraction or repulsion[73–75].

After exploring how hyaluronan behaved as a single-component system, we proceeded to analyze the diffusive behavior emerging from the interaction of hyaluronan with collagen fibrils, the main component of the extracellular matrix in cartilage and many other connective tissues. Previous studies suggested that collagen pose a major hindrance with respect to diffusion[76,77], although these studies were executed with nanosized tracer particles, much smaller than the ones we investigated. While particles in both of the one-component networks exhibited simple diffusive dynamics, where the intermediate scattering function measured with DDM relaxed as a single exponential, they exhibited caged dynamics in the composite networks, as evidenced by a plateau in the ISF and by subdiffusive exponents lower than 1. While the overall structure of the collagen matrix was not visibly affected by the presence of hyaluronan, we cannot exclude that electrostatic interactions might induce spatial heterogeneities in hyaluronan[73,78] which are too small to be detected by light microscopy. We notice that we do see subdiffusive behavior when we increase the collagen concentration from 1 to 2 mg/ml. At this higher collagen concentration, the ISF does not fully relax and the decay of τ(q) deviates from purely diffusive behavior. At high collagen concentration, the large rigidity of the collagen fibers likely contributes a caging effect, consistent with a previous study on composite networks of semiflexible actin and rigid microtubules[46]. In the composite network, the confining agent (hyaluronan) is much softer than collagen, but the mesh size is much smaller. Interestingly, we do not observe subdiffusive motion in a pure hyaluronan system at 4 mg/mL, suggesting the need of an elastically active element to induce subdiffusion. A further possible reason for the emergence of caged dynamics in the composite system is an effective interaction between the components, possibly related to the high negative charge of the hyaluronan and positive charges on collagen[79], which can induce spatial heterogeneities in the hyaluronan network.

## Conclusions

We have shown that, depending on the degree of crosslinking and on the interaction with other components of the extracellular matrix, the diffusion of tracer particles in hyaluronan networks can change drastically. This is interesting when considering physiological situations where physical[80] or chemical[71] crosslinking occurs, as crosslinking is likely to influence the diffusion of nutrients, growth factors and other signaling macromolecules through tissues. While semidilute solutions and transiently crosslinked networks of hyaluronan hinder particle transport through enhanced viscous drag, permanently crosslinked networks mainly exert a size exclusion effect whereby particle mobility is dependent on its size. We furthermore revealed that composite networks of hyaluronan and collagen more strongly restrict particle mobility than expected from the sum of the two parts. This effect is likely related to the large rigidity of collagen fibers, since measurements on collagen-only networks of enhanced density also show caged dynamics. These observations echo recent findings in composite networks of semiflexible actin filaments and more rigid microtubules, where subdiffusion was also observed in response to the presence of rigid filaments[81]. Our results are interesting for understanding how diffusion is affected in the extracellular matrix of tissues, with potential implications in targeting the tumor microenvironment[82] and in the design of hyaluronan-based gels for drug delivery[83] and tissue regeneration[84,85].

## Conflict of interest

There are no conflicts to declare.

## Acknowledgments

We thank B. Mulder (AMOLF) for a critical reading of the manuscript, L. van Buren (AMOLF/TU Delft) for help with the microscopy setup, B.C. Ilochonwu and T. Vermonde (Utrecht University) for advice on chemically crosslinking hyaluronan. The work of F.B., G.P. and G.H.K. was part of the Industrial Partnership Programme Hybrid Soft Materials that is carried out under an agreement between Unilever Research and Development B.V. and the Netherlands Organisation for Scientific Research (NWO). T.S. was supported by a postdoctoral Research Fellowship of the Deutsche Forschungsgemeinschaft (DFG).

# Supplementary Information
## Particle diffusion in extracellular hydrogels

Federica Burla[1], Tatjana Sentjabrskaja[1], Galja Pletikapic[1], Joey van Beugen[1], Gijsje H. Koenderink[1,2,#]

[1] *AMOLF, Department of Living Matter, Biological Soft Matter group, Science Park 104, 1098 XG Amsterdam, the Netherlands*

[2] *Current address: Department of Bionanoscience, Kavli Institute of Nanoscience Delft, Delft University of Technology, Van der Maasweg 9, 2629 HZ Delft, the Netherlands*

[#] *Corresponding author: g.h.koenderink@tudelft.nl*

**List of Supplementary Figures:**

**Figure S1:** Concentration dependence of the shear viscosity of semidilute hyaluronan solutions.

**Figure S2:** DDM measurements of semidilute hyaluronan solutions, comparing high (240 kDa) and low (1.5 MDa) molecular weight samples for 0.6 µm particles.

**Figure S3**: Wavevector dependence of the intermediate scattering function (ISF).

**Figure S4:** Degree of heterogeneity in particle dynamics among ROIs for hyaluronan-only samples for different hyaluronan concentrations and crosslinking states (with 0.6 µm tracer particles).

**Figure S5**: Bulk rheology of hyaluronan solutions and gels in the linear viscoelastic regime.

**Figure S6:** MSD from single particle tracking and τ(q) from DDM for different hyaluronan concentrations and crosslinking states measured with 0.6 µm particles.

**Figure S7:** Subdiffusive exponents α and transport coefficients *K* determined from fitting τ(q), the decay of the ISF, and the lag time dependence of the MSD from particle tracking, for different hyaluronan concentrations and crosslinking states and for 0.6 µm particles.

**Figure S8:** Microrheology of hyaluronan gels at 4 mg/mL from DDM data probed with 0.1, 0.2 and 0.6 µm particles.

**Figure S9:** Rescaling of the intermediate scattering function considering the particle size in crosslinked hyaluronan networks.

**Figure S10:** Subdiffusive exponents α and transport coefficients *K* determined from DDM intermediate scattering functions for tracer particles of different sizes in variously crosslinked hyaluronan networks.

**Figure S11:** Isotropy in x-y displacements for two representative examples of 4 mg/mL semidilute hyaluronan and 2 mg/mL collagen for 0.6 µm particles.

**Figure S12:** Evolution of the width and the non-Gaussian parameter ξ of the van Hove distribution functions with lag time, measured for tracer particles (0.6 µm) in 4 mg/mL hyaluronan networks with varied crosslinking conditions.

**Figure S13:** Microrheology analysis of DDM data for collagen-only network at 1 and 2 mg/mL, semidiluted hyaluronan at 2 mg/mL and for collagen (1 mg/ml)/hyaluronan (2 mg/ml) composite with 0.6 µm particles.

**Figure S14:** Transport coefficients *K* and subdiffusive exponents $\alpha$ from DDM and from particle tracking for tracer particles (0.6 µm) in 1 mg/mL and 2 mg/mL collagen, 2 mg/mL semidiluted hyaluronan, and composite 1 mg/mL collagen- 2 mg/mL hyaluronan networks.

**Figure S15:** Evolution of exponential tails of the van Hove distributions of collagen-only network at 1 mg/mL, semidiluted hyaluronan at 2 mg/mL and for collagen-hyaluronan composite at 1 mg/mL collagen-2 mg/mL hyaluronan with 0.6 µm particles probe size.

**Figure S16:** Non-Gaussian parameter ξ characterizing van Hove distribution functions for collagen-only network at 1 and 2 mg/mL, hyaluronan at 2 mg/mL and for collagen (1 mg/ml)/hyaluronan (2 mg/ml) composite with 0.6 µm particles.

**Figure S17:** Degree of heterogeneity in particle dynamics across ROIs for collagen-only networks and for collagen-hyaluronan composites.

**Figure S18:** ISF from DDM and van Hove distribution from particle tracking for a collagen-only network at a concentration of 2 mg/mL.

**Figure S19:** Linear rheology of 1 and 2 mg/mL collagen-only and composite 1 mg/mL collagen-2 mg/mL hyaluronan networks.

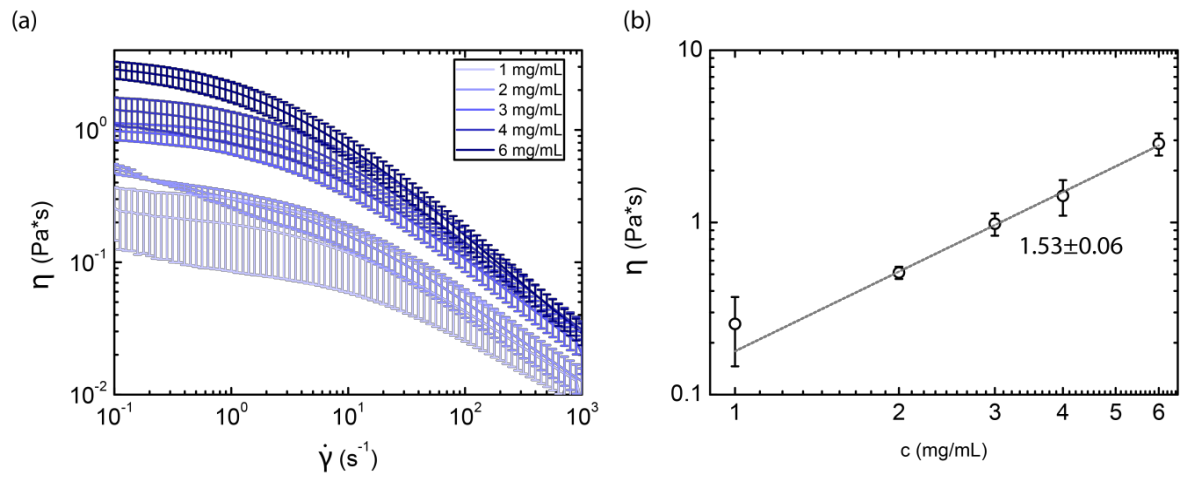

**Supplementary Figure S1: Concentration dependence of the shear viscosity of semidilute hyaluronan solutions.** (a) Shear viscosity as a function of shear rate for semidilute hyaluronan solutions with concentrations from 1 mg/mL to 6 mg/mL (see legend) determined with shear rheometry. (b) The low-shear viscosity (reported for $\dot{\gamma} = 0.1\ s^{-1}$) scales with concentration $c$ with a power law relationship $\eta = ac^b$, where b is 1.53±0.06. This exponent is roughly consistent with the expected b≈1.3 scaling of semidilute polymer solutions. The data represents an average over three independent measurements and the error reported is the standard error of the mean.

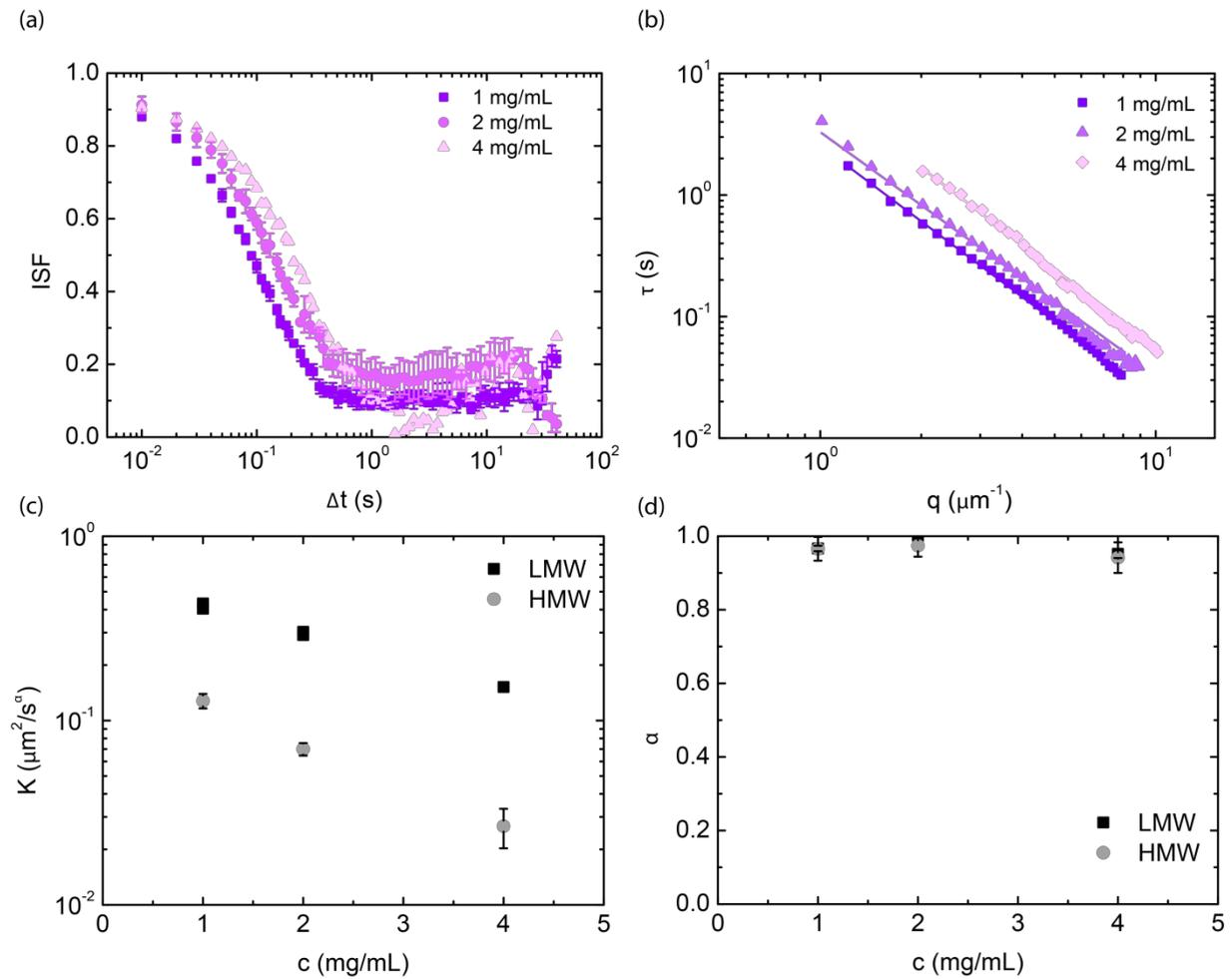

**Supplementary Figure S2: DDM measurements of semidilute hyaluronan solutions, comparing high (240 kDa) and low (1.5 MDa) molecular weight samples for 0.6 μm particles.** (a) ISF for hyaluronan of low molecular weight (240 kDa) at different concentrations, without chemical crosslinking. (b) τ(q) for the same samples. (c) Comparison of transport coefficient $K$ and (d) subdiffusive exponent $\alpha$ for high molecular weight (HMW) and low molecular weight (LMW) samples.

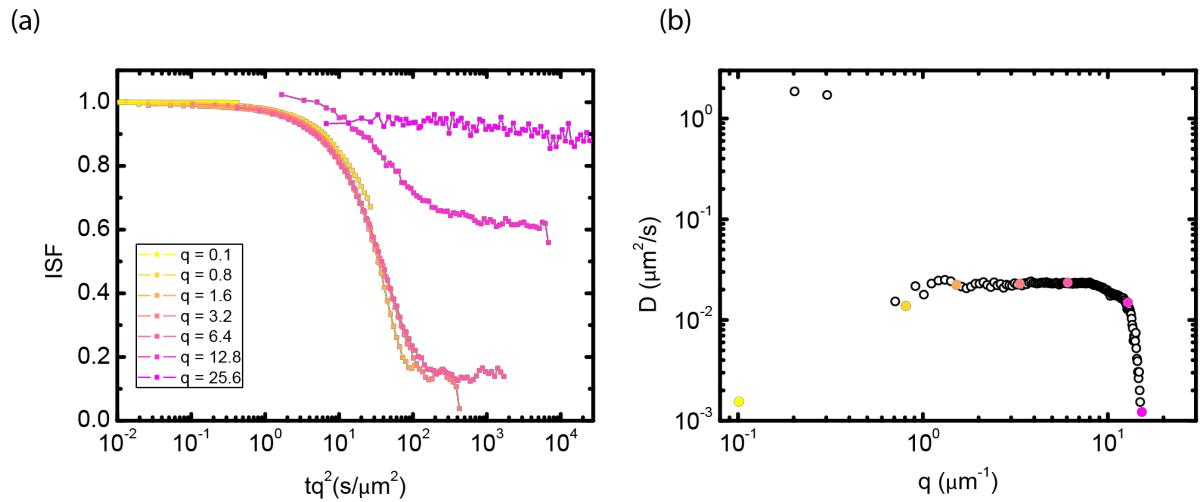

**Supplementary Figure S3: Wavevector dependence of the intermediate scattering function (ISF).** (a) ISF for 0.6 µm tracer particles embedded in 4 mg/mL (semidilute) hyaluronan solutions. Different colours indicate different $q$-values (expressed in the legend in units of µm$^{-1}$). For low $q$-values (0.1 µm$^{-1}$), the ISF does not fully decorrelate, because the measurement spans the whole field of view. At high $q$-values ($\geq$12.8 µm$^{-1}$), the measurement is dominated by noise because of the drifting of particles outside the field of view. (b) Apparent diffusion coefficient obtained from stretched exponential fits of the ISFs as a function of $q$, with colored symbols corresponding to the corresponding curves in panel a.

Particle diffusion in extracellular hydrogels

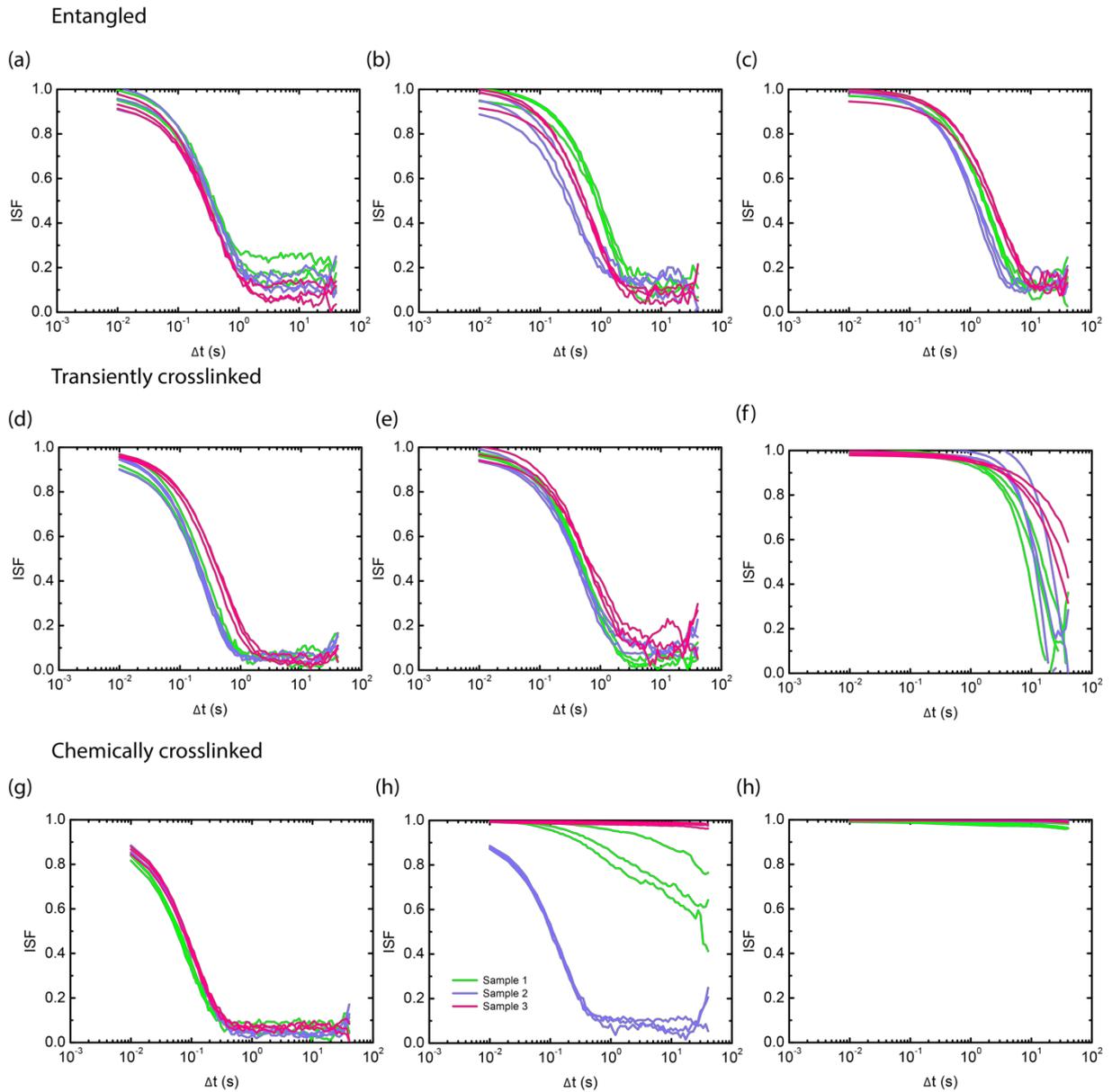

**Supplementary Figure S4: Degree of heterogeneity in particle dynamics among ROIs for hyaluronan-only samples for different hyaluronan concentrations and crosslinking states (with 0.6 µm tracer particles).** (a-b-c) ISF at $q=4.5$ µm$^{-1}$ for semidilute samples, transiently crosslinked samples (d-e-f) and chemically crosslinked samples (g-h-i) at hyaluronan concentrations of 1 mg/mL (left column), 2 mg/mL (central column) and 4 mg/mL (right column). Each color represents an independently prepared sample, and each line represents data from one region of interest.

Particle diffusion in extracellular hydrogels

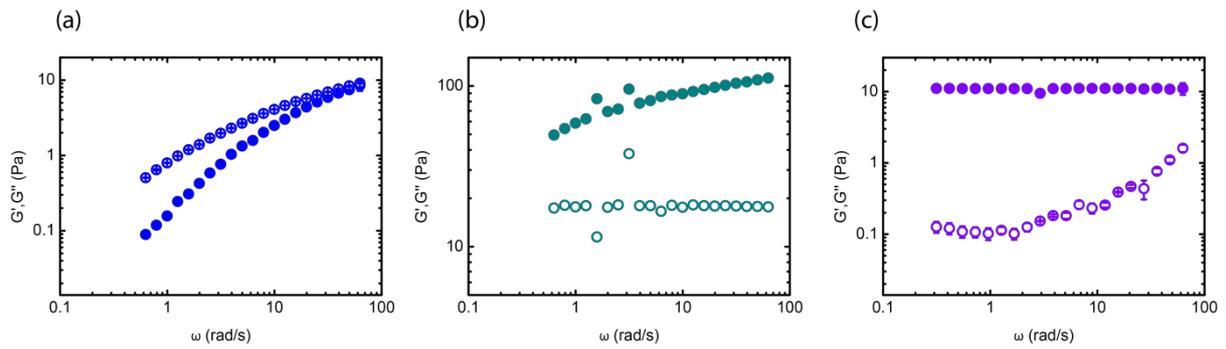

**Supplementary Figure S5: Bulk rheology of the hyaluronan solutions and gels in the linear viscoelastic regime.** (a) Frequency sweep of a semidilute hyaluronan solution at 4 mg/mL, which shows predominantly viscous behavior. (b) Frequency sweep of a transiently crosslinked hyaluronan gel (pH 2.5, putty state) at 4 mg/mL, which behaves as a transient gel. (c) Frequency sweep of a covalently crosslinked hyaluronan gel at 4 mg/mL, which behaves as a soft solid. The results are shown as an average of three repeats over independently prepared samples. The error bar represents the standard error of the mean. Full circles indicate elastic shear moduli $G'$ and empty circles indicate loss shear moduli $G''$.

Particle diffusion in extracellular hydrogels

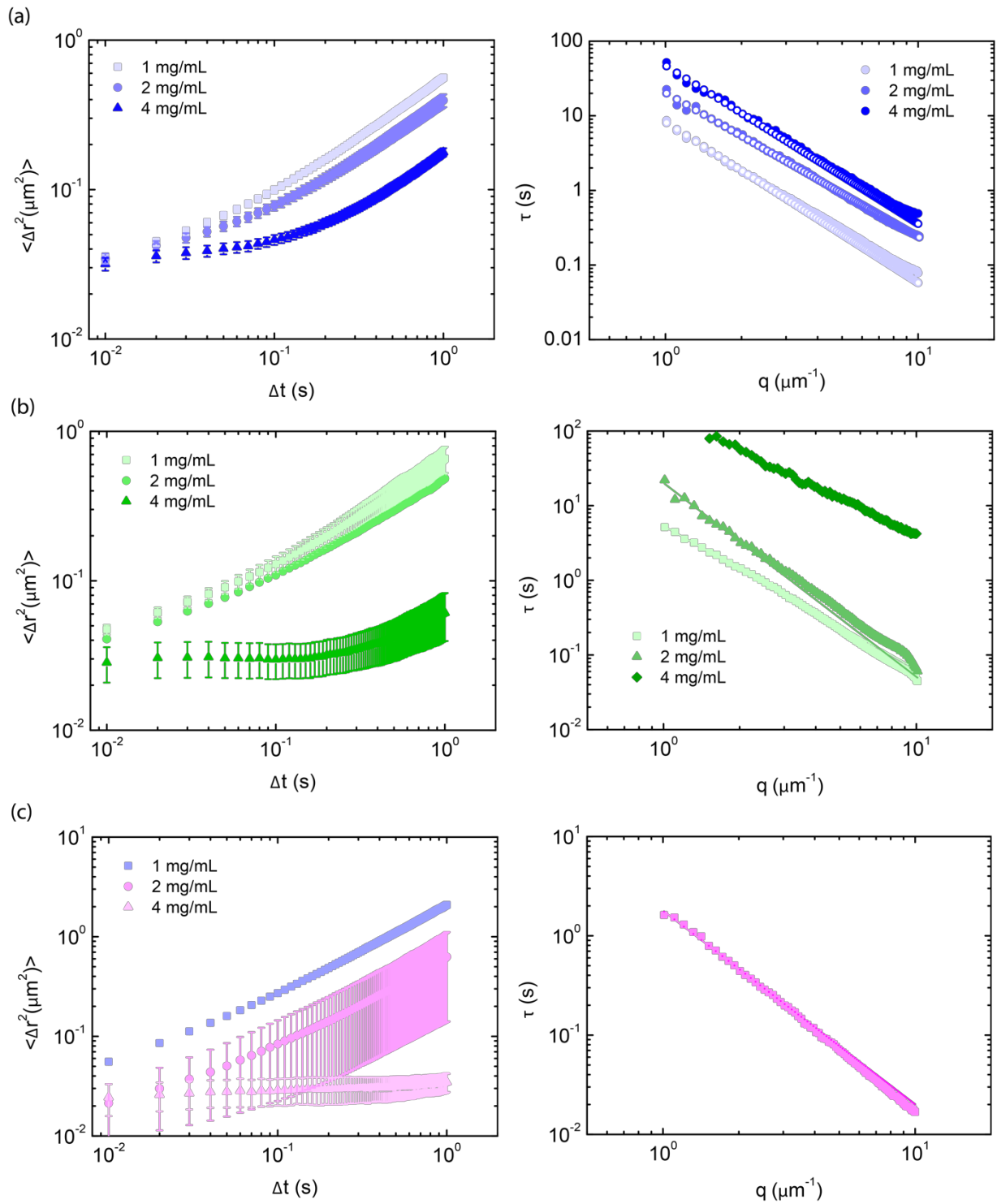

**Supplementary Figure S6: MSD from single particle tracking and τ(q) from DDM for different hyaluronan concentrations and crosslinking states measured with 0.6 µm particles.** Left panels: MSD determined from single-particle tracking for (a) entangled, (b) transiently crosslinked and (c) chemically crosslinked networks at concentrations between 1 and 4 mg/mL. Right panels: relaxation time $\tau$ obtained from fitting ISFs to a stretched exponential, for different values of $q$. In the crosslinked case, we show only the value of $\tau$ at 1 mg/mL because the results at 2 mg/mL and 4 mg/mL do not show a full relaxation for the ISF.

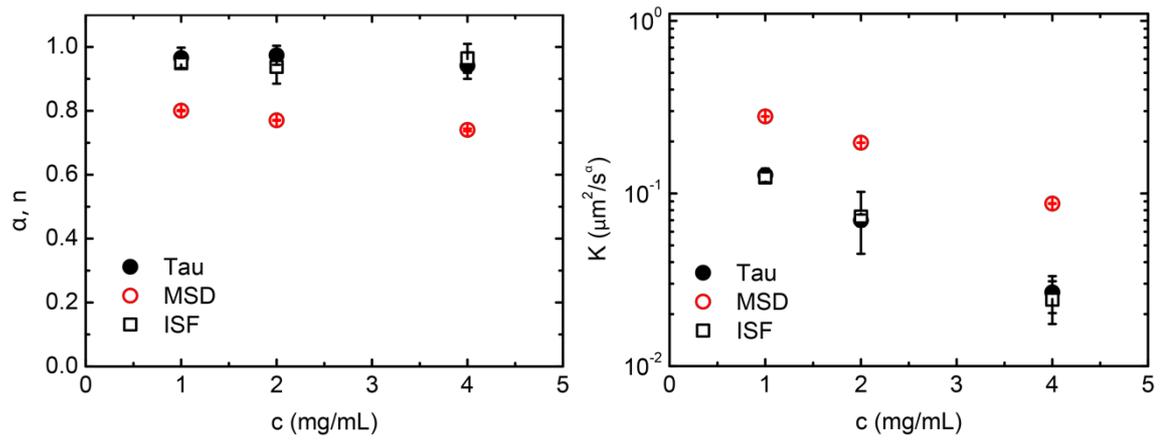
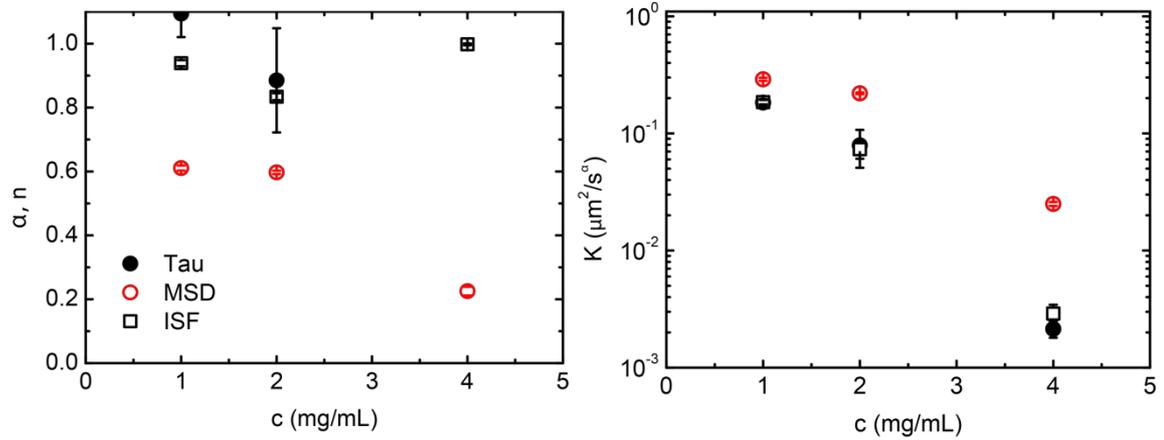
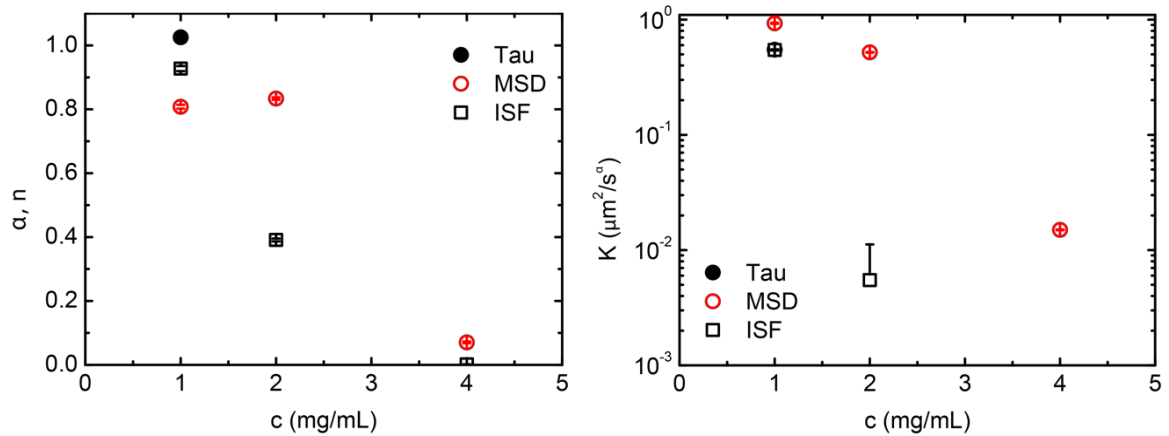

**Supplementary Figure S7: Subdiffusive exponents α and transport coefficients *K* determined from fitting τ(q), the decay of the ISF, and the lag time dependence of the MSD from particle tracking, for different hyaluronan concentrations and crosslinking states and for 0.6 μm particles.** Results show the subdiffusive exponent α or stretching exponent *n* (left panel) and transport coefficient *K* (right panel) for semidilute (a), transiently crosslinked (b) and chemically crosslinked (c) networks. In all cases, the subdiffusive exponents obtained from MSD are below the values obtained from DDM. This is likely related to the short time-interval over which the MSD traces could be obtained.

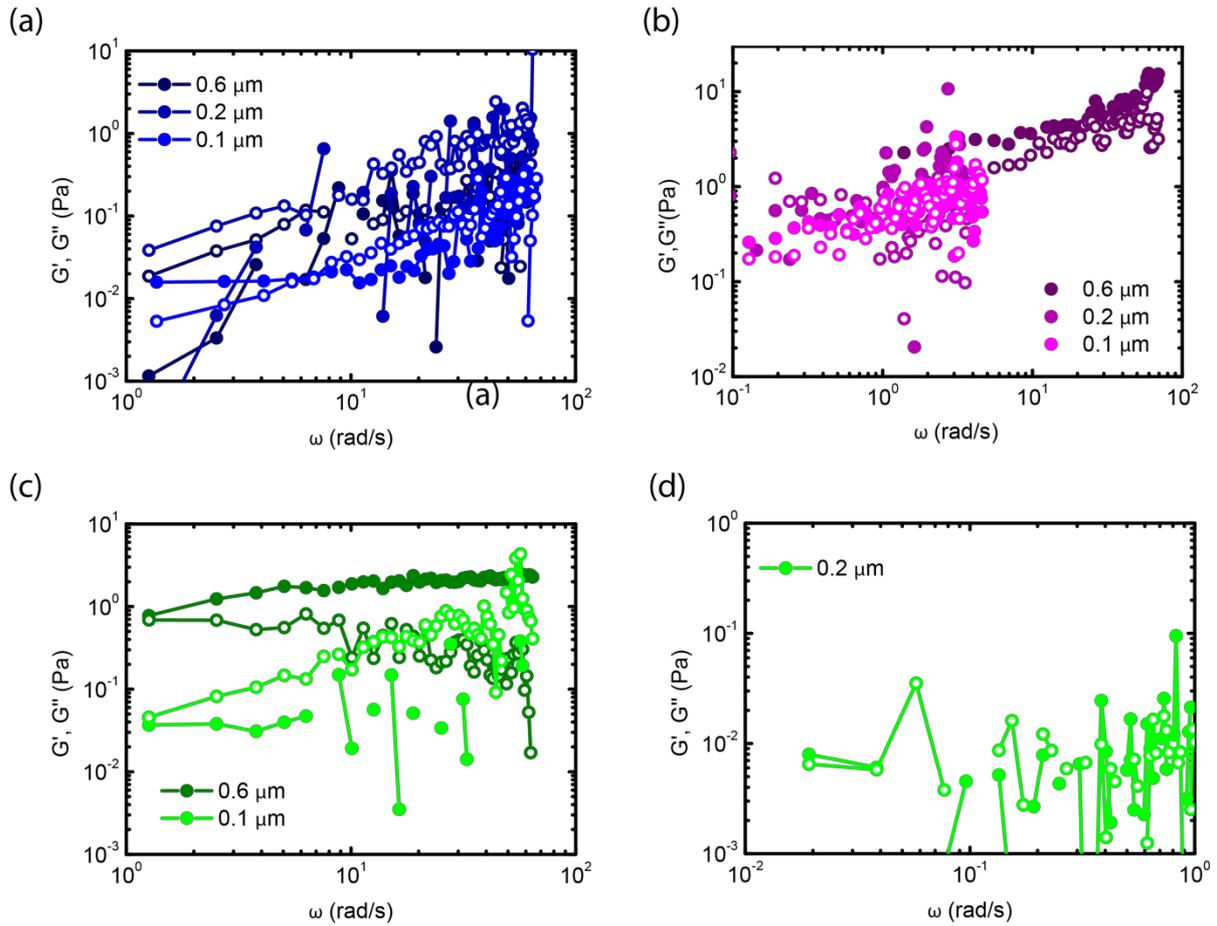

**Supplementary Figure S8: Microrheology of hyaluronan gels at 4 mg/mL from DDM data probed with 0.1 μm, 0.2 μm and 0.6 μm particles.** Microrheology storage and loss moduli for (a) semidilute, (b) permanently crosslinked and (c) transiently crosslinked hyaluronan networks obtained for different particle sizes, obtained from applying the Evans-Tassieri analysis method to the mean-squared-displacement obtained from DDM. Panel (d) represents microrheology of transiently crosslinked networks at a probe particle size of 0.2 μm. Solid circles represent the storage moduli G', and empty circles the loss moduli G''.

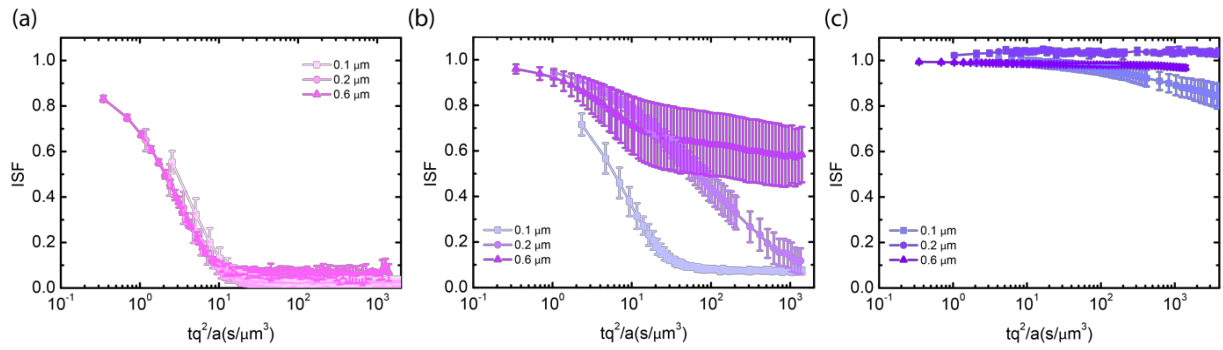

**Supplementary Figure S9: Rescaling of the intermediate scattering function considering the particle size in crosslinked hyaluronan networks.** (a) Collapse of the ISF of a 1 mg/mL crosslinked hyaluronan network for different particle sizes (see legend), obtained by dividing the x-axes for the particle diameter a. (b) Same rescaling for a 2 mg/mL and (c) a 4 mg/mL crosslinked hyaluronan network. Here, the size dependence of the diffusivity deviates from Stokes-Einstein behavior.

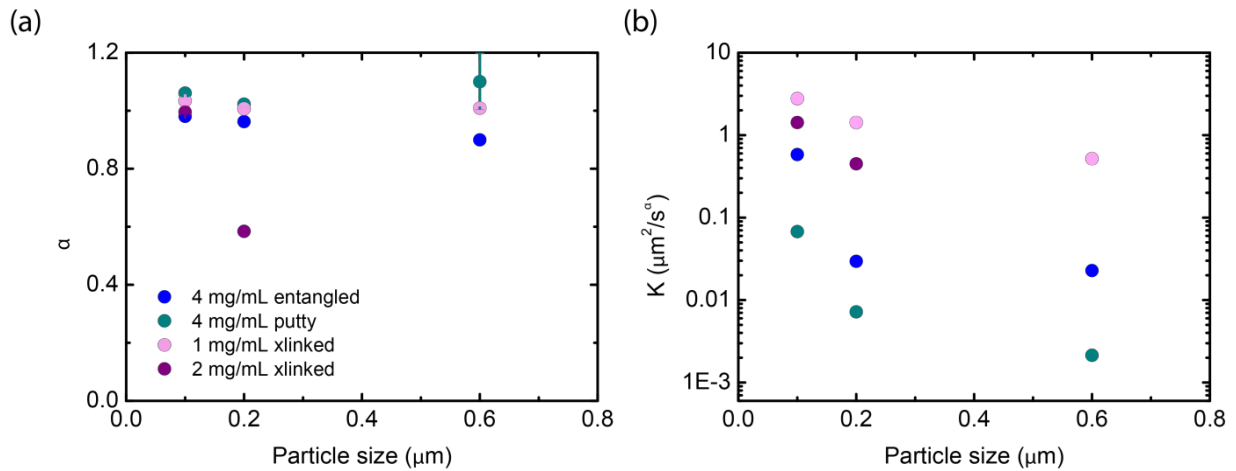

**Supplementary Figure S10: Subdiffusive exponents $\alpha$ and transport coefficients $K$ determined from DDM intermediate scattering functions for tracer particles of different sizes in variously crosslinked hyaluronan networks.** (a) Subdiffusive exponent for different hyaluronan gels (see Legend) as a function of tracer particle diameter, obtained from fitting the decay of τ (q). (b) Transport coefficient for different hyaluronan gels as a function of tracer particle diameter. Values represent averages over at least 9 measurements obtained over three independently prepared samples, and the error shown is the standard error of the mean. For the crosslinked networks, we could not analyze data obtained at 4 mg/mL hyaluronan because of the incomplete decay of the ISF.

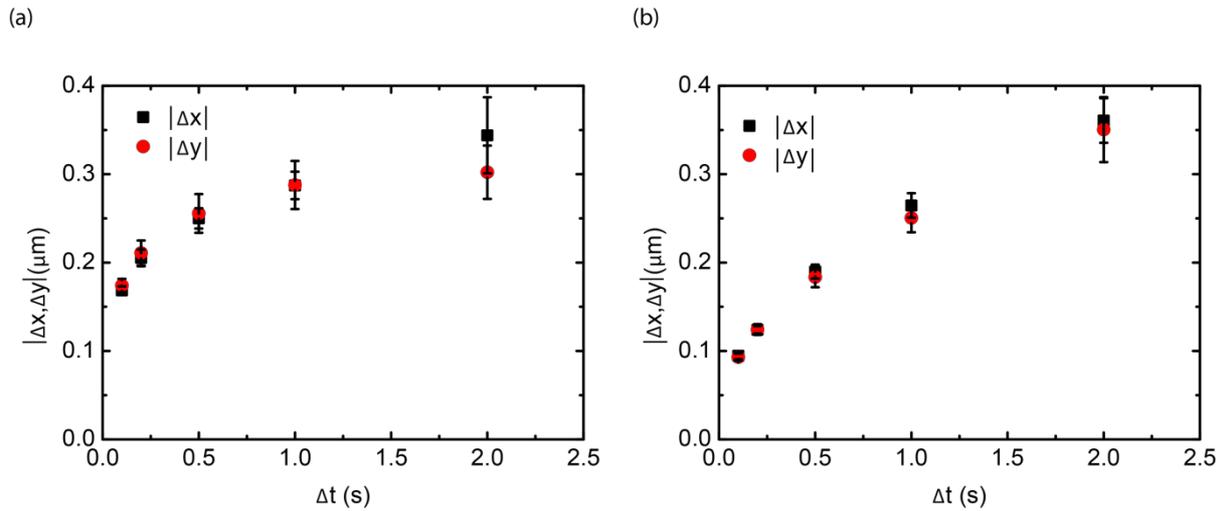

**Supplementary Figure S11: Isotropy in x-y displacements for two representative examples of 4 mg/mL semidilute hyaluronan and 2 mg/mL collagen for 0.6 μm particles.** Representative examples are shown for a (a) collagen-only network at 2 mg/mL and (b) an entangled hyaluronan network at 4 mg/mL for different lag times.

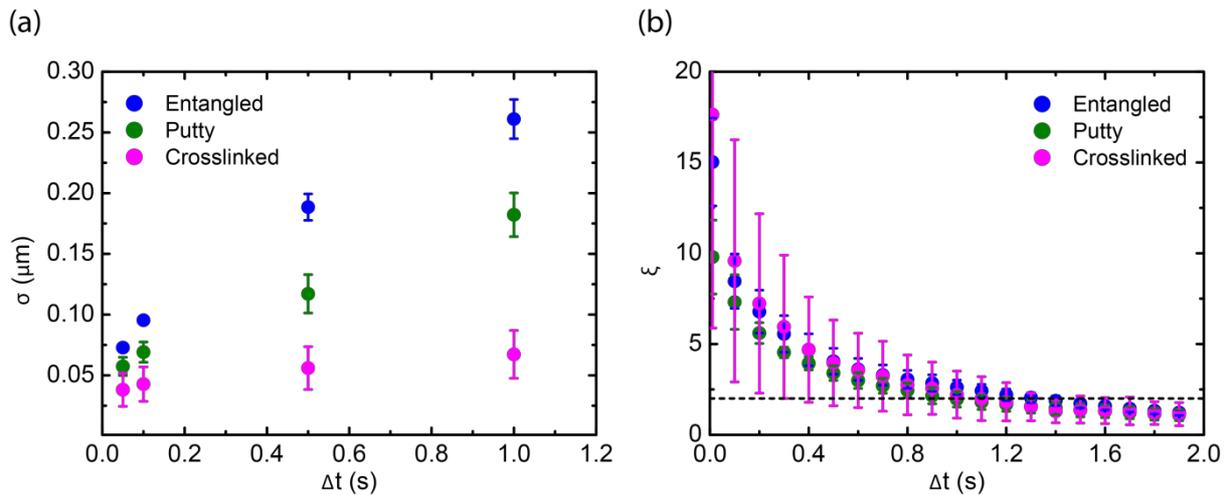

**Supplementary Figure S12: Evolution of the width and the non-Gaussian parameter $\xi$ of the van Hove distribution functions with lag time, measured for tracer particles (0.6 μm) in hyaluronan networks with varied crosslinking conditions.** (a) The width was obtained from fitting the van Hove distributions to a Gaussian function for semidilute solutions (blue), transiently crosslinked gels (green), and chemically crosslinked gels (purple) of hyaluronan (all at 4 mg/ml). Data represents an average over three independent measurements and the errors shown represent the standard error of the mean (b) Non-gaussian parameter calculated as a function of lag time for entangled (blue), transiently crosslinked (green) and chemically crosslinked (pink) samples. The dotted line represents the expected value for an exponential distribution ($\xi$ =2). The monotonic decrease of $\xi$ with lag-time reflects the fact that the displacements are initially smaller than the tracking accuracy (0.1 μm).

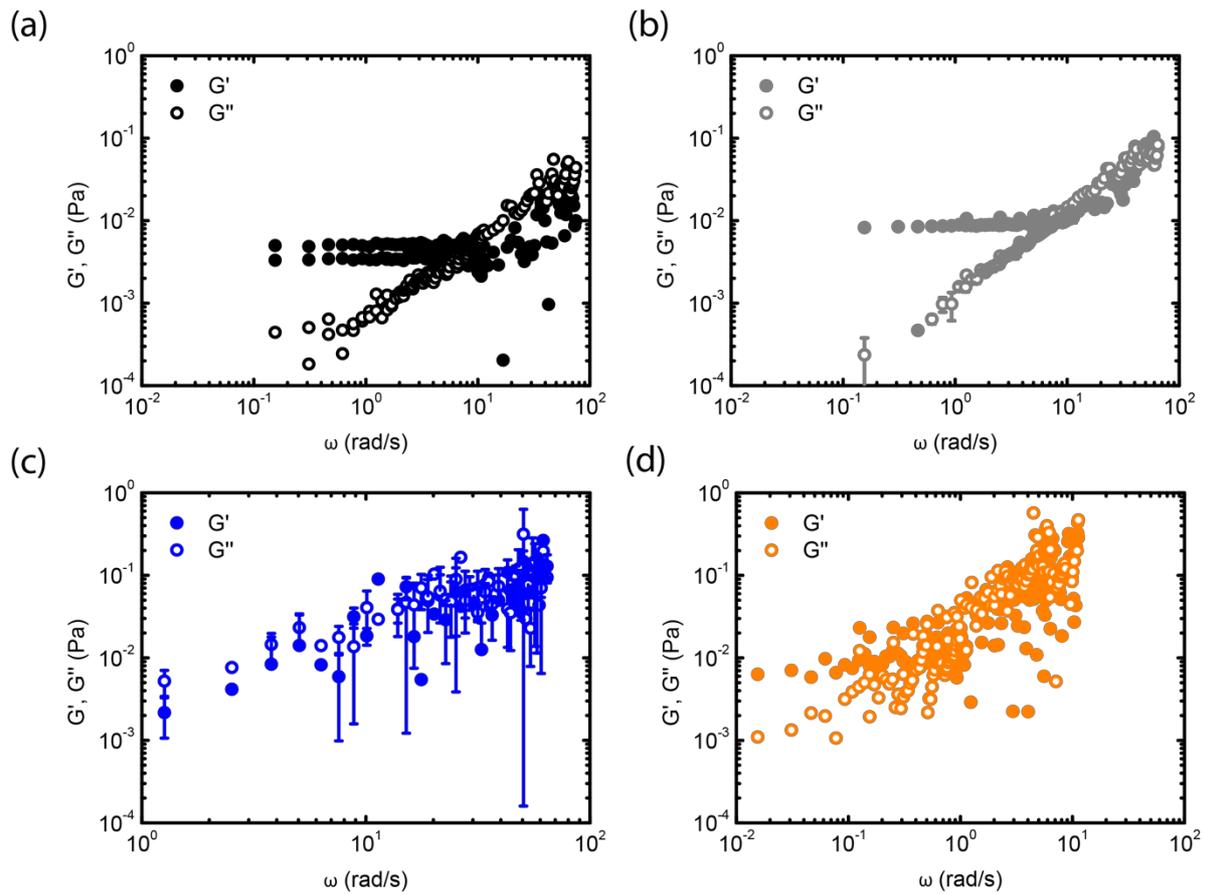

**Supplementary Figure S13: Microrheology analysis of DDM data for collagen-only network at 1 and 2 mg/mL, semidilute hyaluronan at 2 mg/mL, and collagen (1 mg/mL)/hyaluronan (2 mg/mL) composite networks with 0.6 μm particles.** (a) Microrheology for 1 mg/mL collagen, (b) 2 mg/mL collagen, (c) 2 mg/mL pure hyaluronan and (d) composite composed of 2 mg/mL hyaluronan and 1 mg/mL collagen.

Particle diffusion in extracellular hydrogels

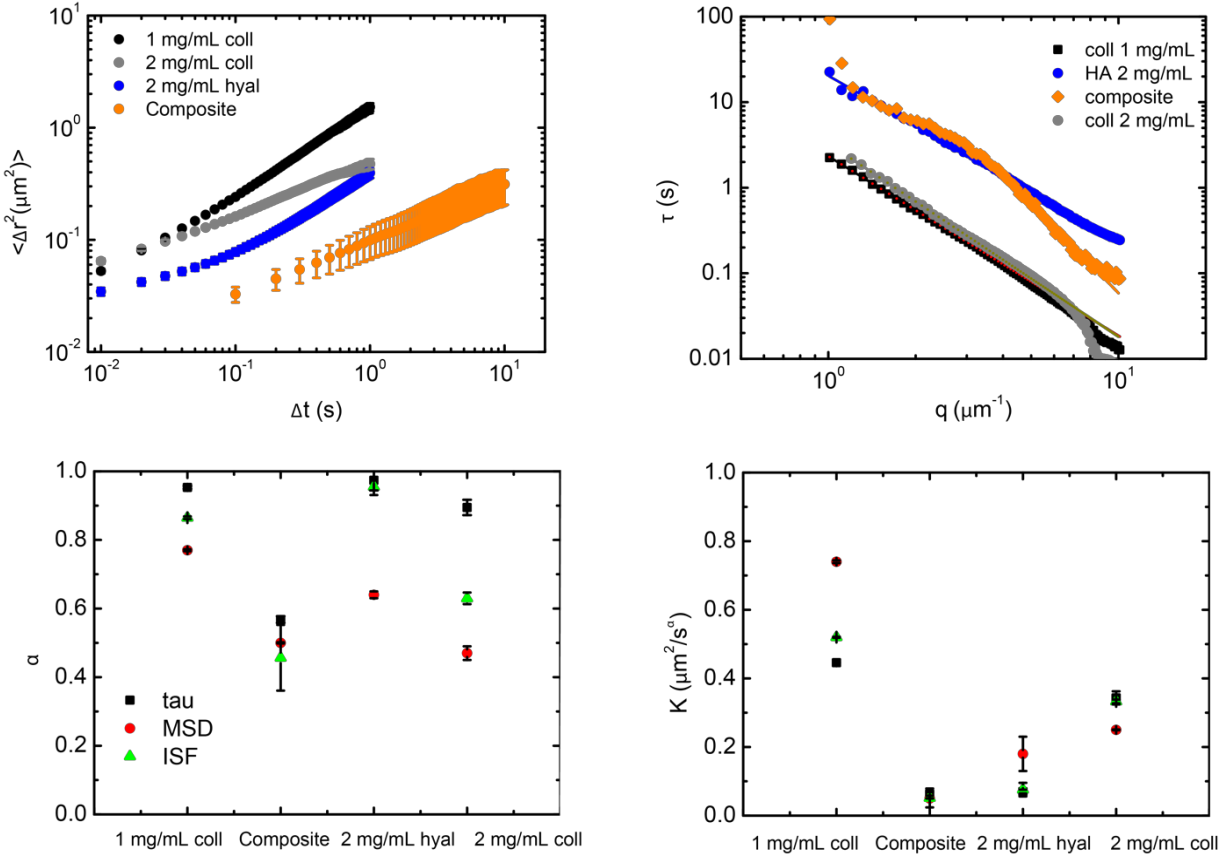

**Supplementary Figure S14: Transport coefficients *K* and subdiffusion exponents α from DDM and from particle tracking for tracer particles (0.6 µm) in 1 mg/mL and 2 mg/mL collagen, 2 mg/mL semidiluted hyaluronan, and composite 1 mg/mL collagen- 2 mg/mL hyaluronan networks.** (a) MSD from particle tracking, (b) τ(*q*) from DDM for composite networks of 1 mg/mL collagen and 2 mg/mL hyaluronan, for collagen-only networks (either 1 or 2 mg/ml), and for 2 mg/mL hyaluronan networks, and (c) subdiffusive exponents and (d) transport coefficients obtained for the same networks.

Particle diffusion in extracellular hydrogels

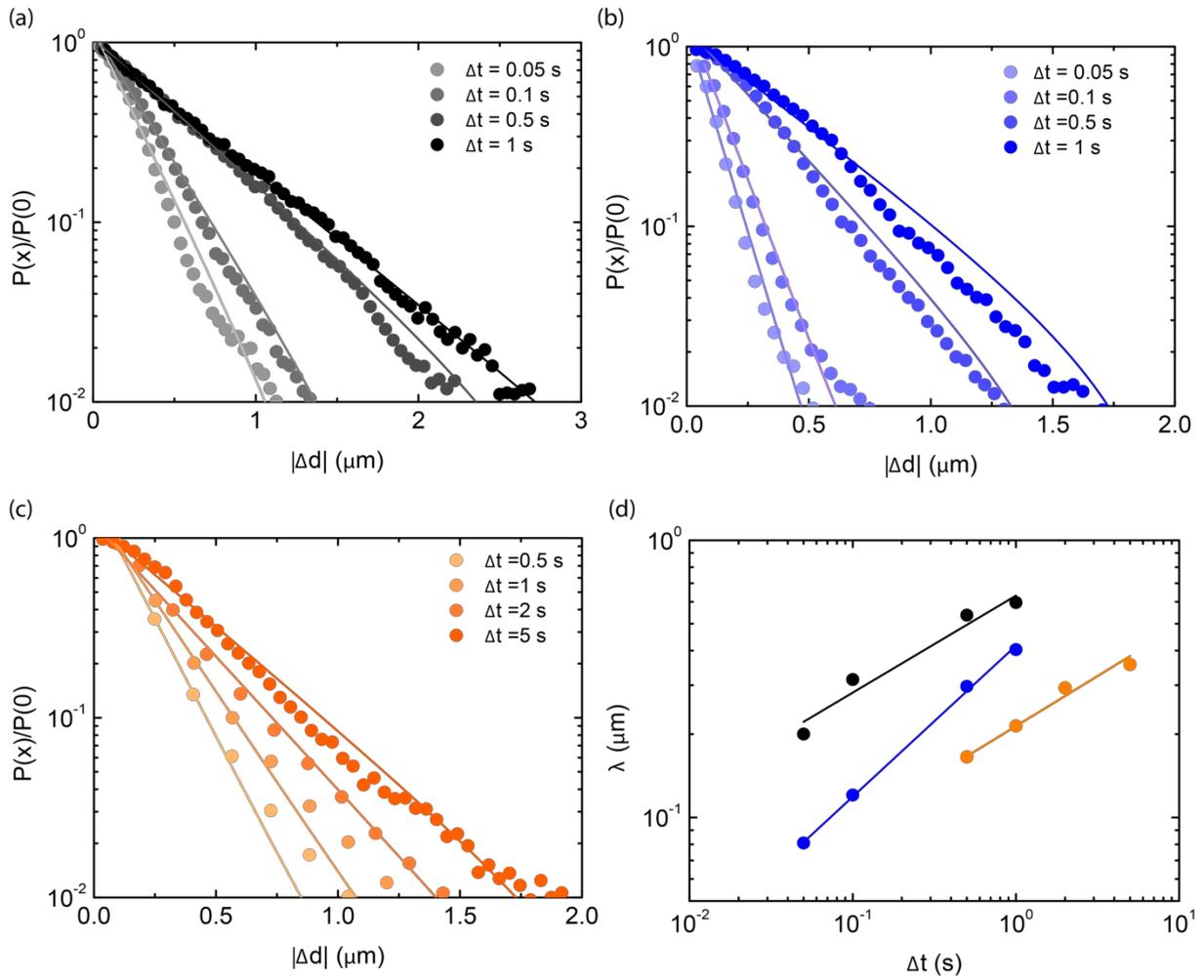

**Supplementary Figure S15: Evolution of exponential tails of the van Hove distributions of collagen-only network at 1 mg/mL, semidiluted hyaluronan at 2 mg/mL and collagen-hyaluronan composite at 1 mg/mL collagen- 2 mg/mL hyaluronan with 0.6 µm particles probe size.** Exponential decay for the van Hove distribution of (a) 1 mg/mL collagen networks, (b) 2 mg/mL hyaluronan solution and (c) 1 mg/mL collagen-2 mg/mL hyaluronan network. (d) Evolution of the decay exponent $\lambda$ a as a function of lag time. Lines represent fits with power law slopes of: 0.35 ± 0.05 (collagen, black), 0.54 ±0.01 (hyaluronan, blue), 0.36 ± 0.03 (composite, orange). The error represents the error from the fit and the data points are obtained by pooling together all of the distributions of the data points for at least three different samples and over at least three different ROIs.

Particle diffusion in extracellular hydrogels

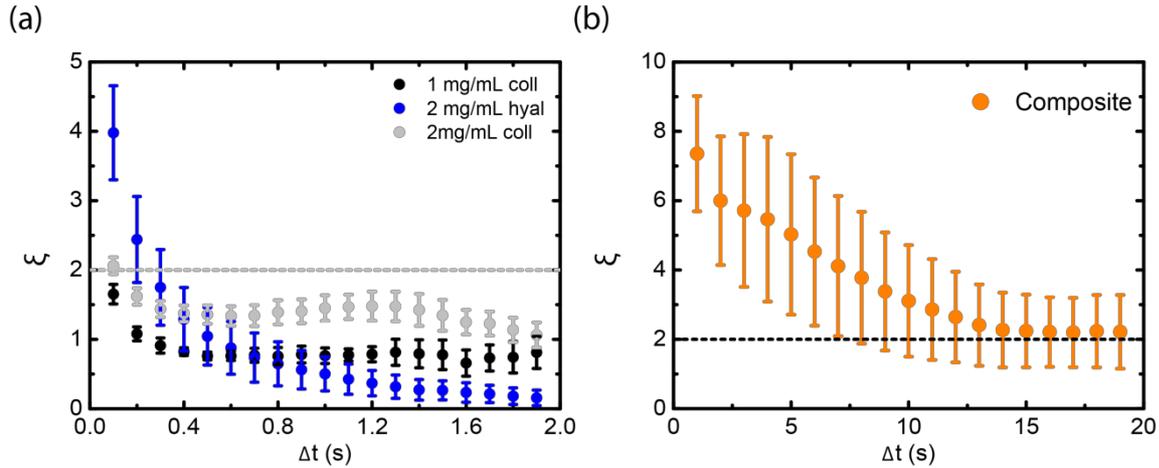

**Supplementary Figure S16: Non-Gaussian parameter $\xi$ characterizing van Hove distribution functions for collagen-only network at 1 and 2 mg/mL, hyaluronan at 2 mg/mL and for collagen (1 mg/ml)/hyaluronan (2 mg/ml) composite with 0.6 μm particles.** (a) Non-Gaussian parameter calculated as a function of lag time for collagen (black), hyaluronan (blue) and composite (orange) samples. The dashed lines indicate the calculated value of $\xi$ for an exponential distribution. The different timescales are related to the different acquisition times used for the samples.

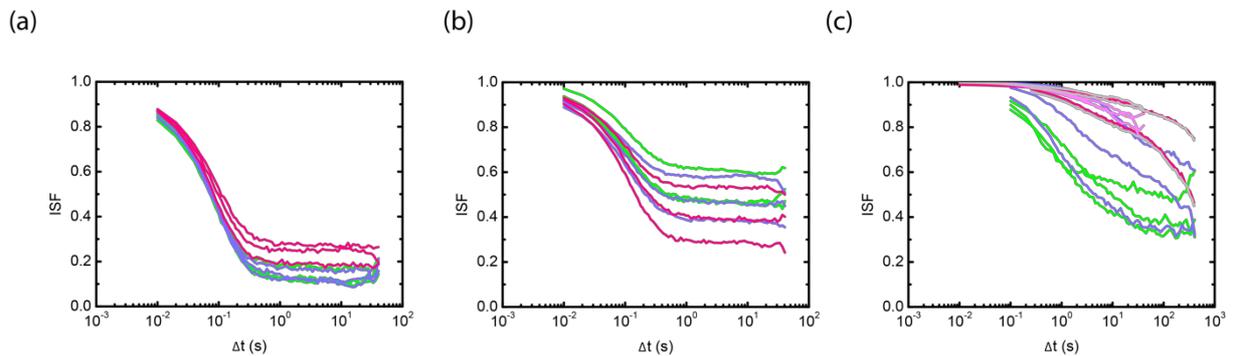

**Supplementary Figure S17: Degree of heterogeneity in particle dynamics across ROIs for collagen-only networks and for collagen-hyaluronan composites.** (a) ISF obtained for $q = 4.5$ μm$^{-1}$ for (a) 1 mg/mL pure collagen network, (b) 2 mg/mL collagen-only network, and (c) a composite of 1 mg/mL collagen and 2 mg/mL hyaluronan. Each color represents an independently prepared sample, and each line represents one ROI.

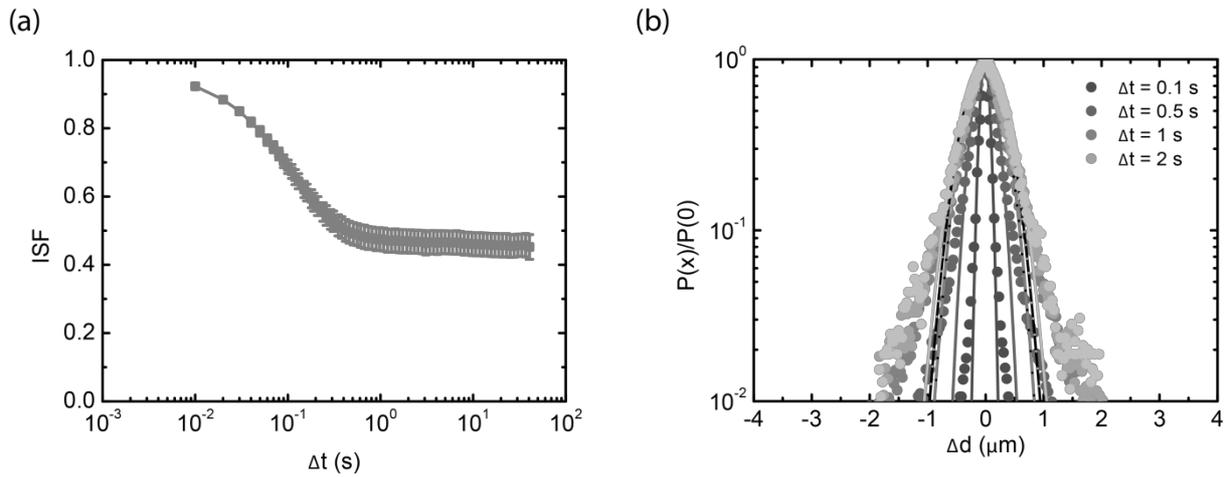

**Supplementary Figure S18: ISF from DDM and van Hove distribution from particle tracking for a collagen-only network at a concentration of 2 mg/mL.** (a) ISF at $q$= 4.5 $\mu m^{-1}$ and (b) van Hove distribution at different lag times.

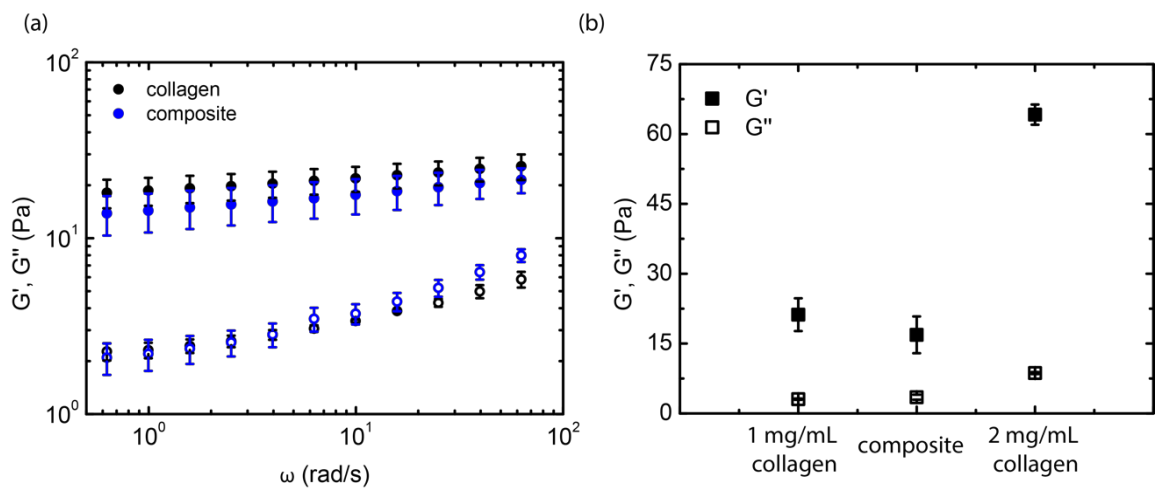

**Supplementary Figure S19: Linear rheology of 1 and 2 mg/mL collagen-only and composite 1 mg/mL collagen-2 mg/mL hyaluronan networks.** (a) Frequency sweep for a collagen network at 1 mg/mL (black symbols) and a composite collagen-hyaluronan network (1 and 2 mg/mL, respectively, blue symbols). The measurement represents an average over at least three different repeats, and the standard deviation represents the standard error of the mean. (b) Elastic (full symbols) and viscous (empty symbols) moduli for collagen networks at 1 and 2 mg/mL and composite networks with 1 mg/mL collagen and 2 mg/mL hyaluronan, measured at a frequency of 6.28 rad/s.

Particle diffusion in extracellular hydrogels